\documentclass[twocolumn,pre,aps,showpacs,showkeys,amsmath]{revtex4}
\usepackage{graphicx}

\begin{document}

\title{Burst Synchronization in A Scale-Free Neuronal Network with Inhibitory Spike-Timing-Dependent Plasticity}
\author{Sang-Yoon Kim}
\email{sykim@icn.re.kr}
\author{Woochang Lim}
\email{wclim@icn.re.kr}
\affiliation{Institute for Computational Neuroscience and Department of Science Education, Daegu National University of Education, Daegu 42411, Korea}

\begin{abstract}
We are concerned about burst synchronization (BS), related to neural information processes in health and disease, in the Barab\'{a}si-Albert scale-free network (SFN) composed of inhibitory bursting Hindmarsh-Rose neurons.
This inhibitory neuronal population has adaptive dynamic synaptic strengths governed by the inhibitory spike-timing-dependent plasticity (iSTDP). In previous works without considering iSTDP, BS was found to appear in a range of noise intensities for fixed synaptic inhibition strengths. In contrast, in our present work, we take into consideration iSTDP and investigate its effect on BS by varying the noise intensity. Our new main result is to find occurrence of a Matthew effect in inhibitory synaptic plasticity: good BS gets better via LTD, while bad BS get worse via LTP. This kind of Matthew effect in inhibitory synaptic plasticity is in contrast to that in excitatory synaptic plasticity where good (bad) synchronization gets better (worse) via LTP (LTD). We note that, due to inhibition, the roles of LTD and LTP in inhibitory synaptic plasticity are reversed in comparison with those in excitatory synaptic plasticity.
Moreover, emergences of LTD and LTP of synaptic inhibition strengths are intensively investigated via a microscopic method based on the distributions of time delays between the pre- and the post-synaptic burst onset times.
Finally, in the presence of iSTDP we investigate the effects of network architecture on BS by varying the symmetric attachment degree $l^*$ and the asymmetry parameter $\Delta l$ in the SFN.
\end{abstract}


\maketitle

\section{Introduction}
\label{sec:INT}
Recently, population synchronization of bursting neurons has attracted much attention in many aspects \cite{BSsync1,BSsync2,BSsync3,BSsync4,BSsync5,BSsync6,BSsync7,BSsync8,BSsync9,BSsync10,BSsync11,BSsync12,BSsync13,BSsync14,BSsync15,
BSsync16,BSsync17,BSsync18,BSsync19,BSsync20,BSsync21,BSsync22,BSsync23,BSsync24,BSsync27,BSsync25,
BSsync26,BSsync28,BSsync29}. There are several representative examples of bursting neurons such as intrinsically bursting neurons and chattering neurons in the cortex \cite{CT1,CT2}, thalamic relay neurons and thalamic reticular neurons in the thalamus \cite{TRN1,TRN2,TR}, hippocampal pyramidal neurons \cite{HP}, Purkinje cells in the cerebellum \cite{PC}, pancreatic $\beta$-cells \cite{PBC1,PBC3,PBC2}, and respiratory neurons in pre-Botzinger complex \cite{BC1,BC2}. As is well known, burstings occur when neuronal activity alternates, on a slow timescale, between a silent phase and an active (bursting) phase of fast repetitive spikings \cite{Rinzel1,Rinzel2,Izhi,Burst1,Burst2,Burst3}. Due to a repeated sequence of spikes in the bursting, there are many hypotheses on the importance of bursting activities in neural computation \cite{Burst5,Burst6,Burst4,Izhi2,Burst2}; for example, (a) bursts are necessary to overcome the synaptic transmission failure, (b) bursts are more reliable than single spikes in evoking responses in post-synaptic neurons, (c) bursts evoke long-term potentiation/depression (and hence affect synaptic plasticity much greater than single spikes), and (d) bursts can be used for selective communication between neurons.

Here, we are interested in burst synchronization (BS) (i.e., synchrony on the slow bursting timescale) which characterizes temporal coherence between the (active phase) burst onset times (i.e., times at which burstings start in active phases). This type of BS is associated with neural information processes in health and disease. For example, large-scale BS (called the sleep spindle oscillation of 7-14 Hz) has been found to occur through interaction between the excitatory thalamic relay cells and the inhibitory thalamic reticular neurons in the thalamus during the early stage of slow-wave
sleep \cite{Spindle1,Spindle2}. These sleep spindle oscillations are involved in memory consolidation \cite{Spindle3,Spindle4}. On the other hand, BS is also correlated to abnormal
pathological rhythms, associated with neural diseases such as movement disorder (Parkinson's disease and essential tremor) \cite{PD1,PD5,PD2,PD4,PD3} and epileptic seizure \cite{Epilepsy,PD5}. For the case of the Parkinson's disease hypokinetic motor symptoms (i.e., slowness and rigidity of voluntary movement) are closely associated with BS occurring in the beta band of 10-30 Hz
range in the basal ganglia, while the hyperkinetic motor symptom (i.e., resting tremor) is related to BS in the theta band of 3-10 Hz.

In real brains, synaptic strengths may change for adaptation to the environment (i.e., they can be potentiated \cite{LTP2,LTP1,LTP3} or depressed \cite{LTD1,LTD2,LTD3,LTD4}). These adjustments of synapses are called the synaptic plasticity which provides the basis for learning, memory, and development \cite{Abbott1}. Such synaptic plasticity is taken into consideration in the present work. As to the synaptic plasticity, we consider a spike-timing-dependent plasticity (STDP) \cite{STDP1,STDP2,STDP3,STDP4,STDP5,STDP6,STDP7,STDP8}. For the STDP, the synaptic strengths vary via an update rule depending on the relative time difference between the pre- and the post-synaptic burst onset times. Many models for STDP have been employed to explain results on synaptic modifications occurring in diverse neuroscience topics for health and disease such as temporal sequence learning \cite{TSLearning}, temporal pattern recognition \cite{EtoE6}, coincidence detection \cite{EtoE0}, navigation \cite{Navi}, direction selectivity \cite{DirSel}, memory consolidation \cite{Memory}, competitive/selective development \cite{Devel}, and deep brain stimulation \cite{Lou,Minesota}. Recently, the effects of STDP on population synchronization in ensembles of coupled neurons were also studied in various aspects \cite{Tass1,Tass2,Brazil1,Brazil2,SSS,SBS,SSSSTDP,FSSiSTDP}.

Here, we study emergence of BS in a scale-free network (SFN) of inhibitory bursting neurons. In the absence of synaptic plasticity (i.e., coupling strengths are static), BS has been found to appear in a range of noise intensity $D$ for a fixed coupling strength \cite{NN-SFN}. As $D$ is increased from 0, the degree of BS decreases due to a destructive role of noise to spoil the BS, and when passing a threshold a transition from BS to desynchronization occurs. In contrast to the previous work, we now take into consideration the synaptic plasticity, and then the inhibitory population has adaptive dynamic synaptic strengths governed by the inhibitory spike-timing-dependent plasticity (iSTDP). Studies of synaptic plasticity have been mainly focused on excitatory-to-excitatory (E-to-E) synaptic connections between excitatory pyramidal cells \cite{EtoE0,EtoE1,EtoE3,EtoE4,EtoE2,EtoE5,EtoE6,EtoE7,EtoE8}. An asymmetric Hebbian time window was employed for the excitatory STDP (eSTDP) update rule \cite{STDP1,STDP2,STDP3,STDP4,STDP5,STDP6,STDP7,STDP8}. When a pre-synaptic spiking precedes a post-synaptic spiking, long-term potentiation (LTP) occurs; otherwise, long-term depression (LTD) appears. On the other hand, plasticity of inhibitory connections has attracted less attention mainly due to experimental obstacles and diversity of inhibitory interneurons \cite{iSTDP4,iSTDP3,iSTDP2,iSTDP1,iSTDP5}. Along with the advent of fluorescent labeling and optical manipulation of neurons according to their genetic types \cite{iExpM1,iExpM2}, inhibitory plasticity has also begun to be focused. Particularly, studies on iSTDP of inhibitory-to-excitatory (I to E) connections  have been made. Thus, iSTDP has been found to be diverse and cell-specific \cite{iSTDP12,iSTDP4,iSTDP11,iSTDP10,iSTDP8,iSTDP3,iSTDP6,iSTDP7,iSTDP2,iSTDP1,iSTDP5,iSTDP9}.

In this paper, we consider an inhibitory Barab\'{a}si-Albert SFN consisting of bursting neurons \cite{BA1,BA2}, and investigate the effect of iSTDP [of inhibitory-to-inhibitory (I to I) connections] on BS by varying the noise intensity $D$.
As mentioned above, previous studies on iSTDP have been focused mainly on the case of inhibitory-to-excitatory (I to E) connections. Even in this case of I to E iSTDP, time windows for the iSTDP rule vary depending on the target pyramidal cells (e.g., delayed Hebbian time window \cite{iSTDP10,iSTDP8} for the pyramidal cells in the entorhinal cortex and symmetric time window \cite{iSTDP11} for the pyramidal cells in the CA1 hippocampus). In our present work, we consider the I to I iSTDP; the target neuron is inhibitory, in contrast to the above excitatory case. Recently, such I to I iSTDP has been studied in works \cite{Tass1,Lou,Minesota} where the anti-Hebbian time window was used for the I to I iSTDP. Following them, we also employ the anti-Hebbian time window for studying the effect of I to I iSTDP on BS.
As the time is increased, strengths of synaptic inhibition $\{ J_{ij} \}$ are changed, and eventually, they approach saturated limit values after a sufficiently long time.
Depending on $D$, mean values of saturated synaptic inhibition strengths $\{ J_{ij}^* \}$ are potentiated [long-term potentiation (LTP)] or depressed [long-term depression (LTD)], in comparison
with the initial mean value of synaptic inhibition strengths. In contrast, standard deviations from the mean values of LTP/LTD are much increased, when compared with the initial dispersion,
independently of $D$. Both the mean value and the standard deviation (for the distribution of synaptic inhibition strengths) may affect BS.
The LTD (LTP) tends to increase (decrease) the degree of BS due to decrease (increase) in the mean value of synaptic inhibition strengths, and
the increased standard deviations have a tendency to decrease the degree of BS. In most range of $D$ with LTD, good BS (with higher bursting measure) gets better because the effect of mean LTD is dominant in comparison with the effect of increased standard deviation. In contrast, in the range of $D$ with LTP, bad BS (with lower bursting measure) gets worse due to the effects of both LTP and increased standard deviation.
We note that this effect is similar to the Matthew effect in the sociology of science \cite{Matthew}; the rich get richer and the poor get poorer. Hence, for our case, we call the effect of iSTDP on BS as the Matthew effect in inhibitory synaptic plasticity \cite{FSSiSTDP}, in addition to the Matthew effect in excitatory synaptic plasticity in previous works \cite{SSS,SBS,SSSSTDP}.
This type of Matthew effect in inhibitory synaptic plasticity is in contrast to that in excitatory synaptic plasticity where good (bad) synchronization gets better (worse) via LTP (LTD) \cite{SSS,SBS,SSSSTDP}.
We note that the role of LTD (LTP) in the case of iSTDP is similar to that of LTP (LTD) for the case of eSTDP. Emergences of LTD and LTP of synaptic inhibition strengths are also investigated through a microscopic method based on the distributions of time delays between the nearest burst onset times of the pre- and the post-synaptic neurons.  Furthermore, in the presence of iSTDP we study the effects of network architecture on BS for a fixed value of $D$ by varying the symmetric attachment degree $l^*$ and the asymmetry parameter $\Delta l$ in the SFN. Like the above case of variation in $D$, Matthew effects in inhibitory synaptic plasticity are also found to occur by varying $l^*$ and $\Delta l$.

For the sake of clearness, we also make a brief summary of new main results in our present work. In a previous work \cite{NN-SFN}, BS was found to occur in a range of noise intensities for static synaptic inhibition strengths.
However, in real brains, synaptic strengths may change for adaptation to the environment (i.e., they may be potentiated or depressed); these adjustments of synapses are called the synaptic plasticity. In our present work, we take into consideration the iSTDP and investigate its effect on BS by varying the noise intensity, in contrast to the previous works without considering iSTDP. Our new main finding is occurrence of the Matthew effect in inhibitory synaptic plasticity: good BS gets better via LTD, while bad BS get worse via LTP. This type of Matthew effect in inhibitory synaptic plasticity is in contrast to the Matthew effect in excitatory synaptic plasticity \cite{SSS,SBS,SSSSTDP} where good (bad) synchronization gets better (worse) via LTP (LTD). We note that, due to inhibition, the roles of LTD and LTP in inhibitory synaptic plasticity are reversed in comparison with those in excitatory synaptic plasticity. Furthermore, our results on the effect of iSTDP on BS are also expected to be useful for understanding the basis for not only the fundamental brain function (e.g., learning, memory, and development) \cite{TSLearning,Memory,Devel}, but also neural diseases (e.g., Parkinson’s disease and epilepsy) \cite{PD1,Epilepsy,PD5,PD2,PD4,PD3}.

This paper is organized as follows. In Sec.~\ref{sec:SFN}, we describe an inhibitory Barab\'{a}si-Albert SFN of bursting neurons with inhibitory synaptic plasticity.
Then, in Sec.~\ref{sec:BS} we investigate the effects of iSTDP on BS. Finally, a summary is given in Sec.~\ref{sec:SUM}.

\section{Scale-Free Network of Inhibitory Bursting Hindmarsh-Rose Neurons}
\label{sec:SFN}
Synaptic connectivity in brain networks has been found to have complex topology which is neither regular nor completely random \cite{CN6,Buz2,CN1,CN2,CN7,CN3,CN4,CN5,Sporns}.
Particularly, brain networks have been found to exhibit power-law degree distributions (i.e., scale-free property) in the rat hippocampal networks \cite{SF4,SF1,SF2,SF3} and the human cortical functional network \cite{SF5}. Furthermore, robustness against simulated lesions of mammalian cortical anatomical networks \cite{SF9,SF6,SF7,SF8,SF10,SF11} has also been found to be most similar to that of an SFN \cite{SF12}.
Many recent works on various subjects of neurodynamics (e.g., coupling-induced BS, delay-induced BS, and suppression of BS) have been done in SFNs with a few percent of
hub neurons with an exceptionally large number of connections \cite{BSsync11,BSsync12,BSsync14,BSsync15,BSsync19,BSsync26}.

We consider an inhibitory SFN composed of $N$ bursting neurons equidistantly placed on a one-dimensional ring of radius $N/ 2 \pi$. We employ a directed Barab\'{a}si-Albert SFN model (i.e. growth and preferential directed attachment) \cite{BA1,BA2}. At each discrete time $t,$ a new node is added, and it has $l_{in}$ incoming (afferent) edges and $l_{out}$ outgoing (efferent) edges via preferential attachments with $l_{in}$ (pre-existing) source nodes and $l_{out}$ (pre-existing) target nodes, respectively. The (pre-existing) source and target nodes $i$ (which are connected to the new node) are preferentially chosen depending on their out-degrees $d_i^{(out)}$ and in-degrees $d_i^{(in)}$ according to the attachment probabilities $\Pi_{source}(d_i^{(out)})$ and $\Pi_{target}(d_i^{(in)})$, respectively:
\begin{eqnarray}
\Pi_{source}(d_i^{(out)}) &=& \frac{d_i^{(out)}}{\sum_{j=1}^{N_{t -1}}d_j^{(out)}}\;\; \textrm{and} \nonumber \\
\Pi_{target}(d_i^{(in)}) &=& \frac{d_i^{(in)}}{\sum_{j=1}^{N_{t -1}}d_j^{(in)}},
\label{eq:AP}
\end{eqnarray}
where $N_{t-1}$ is the number of nodes at the time step $t-1$.
Hereafter, the cases of $l_{in} = l_{out} (\equiv l^*)$  and $l_{in} \neq l_{out}$ will be referred to as symmetric and asymmetric preferential attachments, respectively.
For generation of an SFN with $N$ nodes, we begin with the initial network at $t=0$, consisting of $N_0=50$ nodes where the node 1 is connected bidirectionally to all the other nodes, but the remaining nodes (except the node 1) are sparsely and randomly connected with a low probability $p=0.1$. The processes of growth and preferential attachment are repeated until the total number of nodes becomes $N$. For this case, the node 1 will be grown as the head hub with the highest degree. Then, the grown network via the above process becomes scale free, because the distributions for the in- and the out-degrees $d^{(in)}$ and $d^{(out)}$ exhibit power-law decays with the same exponent $\gamma=3$, $P(d^{(in)}) \sim {d^{(in)}}^{-\gamma}$ and $P(d^{(out)}) \sim {d^{(out)}}^{-\gamma}$ \cite{BA1,BA2}.

\begin{table}
\caption{Parameter values used in our computations.}
\label{tab:Parm}
\begin{ruledtabular}
\begin{tabular}{llllll}
(1) & \multicolumn{5}{l}{Single HR Bursting Neurons \cite{Longtin}} \\
&  $a=1$ & $b=3$ & $c=1$ & $d=5$ & $r=0.001$ \\
&  $s=4$ & $x_0 = -1.6$ & & &  \\
\hline
(2) & \multicolumn{5}{l}{External Stimulus to HR Bursting Neurons} \\
& \multicolumn{2}{l}{$I_{DC,i} \in [1.3, 1.4]$} & \multicolumn{3}{l}{$D$: Varying} \\
\hline
(3) & \multicolumn{5}{l}{Inhibitory Synapse Mediated by The GABA$_{\rm A}$ } \\
& \multicolumn{5}{l}{Neurotransmitter \cite{GABA}} \\
& $\tau_l=1$ & $\tau_r=0.5$ & $\tau_d=5$ & \multicolumn{2}{l}{$X_{syn}=-2$} \\
\hline
(4) & \multicolumn{5}{l}{Synaptic Connections between Neurons in The} \\
& \multicolumn{5}{l}{Barab\'{a}si-Albert SFN} \\
& \multicolumn{5}{l}{$l^*$: Varying (symmetric preferential attachment)} \\
& \multicolumn{5}{l}{$\Delta l$: Varying (asymmetric preferential attachment)} \\
& $J_0 = 12 $ & $\sigma_0 = 0.1$ & \multicolumn{3}{l}{$J_{ij} \in [0.0001, 20]$} \\
\hline
(5) & \multicolumn{5}{l}{Anti-Hebbian STDP Rule} \\
& $\delta = 0.08$ & $A_{+} = 1.0$ & $A_{-} = 1.3$ & $\tau_{+} = 410$ & $\tau_{-} = 330$ \\
\end{tabular}
\end{ruledtabular}
\end{table}

As an element in our SFN, we choose the representative bursting HR neuron model which was originally introduced to describe the time evolution of the membrane potential for the pond snails \cite{HR1,HR2,HR3}; this HR neuron model was studied in many aspects \cite{BSsync22,BSsync27,BSsync24,HR4,Kim1,Kim2,NN-SFN,BSsync28,BSsync29,HR5}.
We consider the Barab\'{a}si-Albert SFN composed of $N$ HR bursting neurons. The following equations (\ref{eq:PD1})-(\ref{eq:PD3}) govern the population dynamics in the SFN:
\begin{eqnarray}
\frac{dx_i}{dt} &=& y_i - a x^{3}_{i} + b x^{2}_{i} - z_i +I_{DC,i} +D \xi_{i} -I_{syn,i}, \label{eq:PD1} \\
\frac{dy_i}{dt} &=& c - d x^{2}_{i} - y_i, \label{eq:PD2} \\
\frac{dz_i}{dt} &=& r \left[ s (x_i - x_o) - z_i \right], \label{eq:PD3}
\end{eqnarray}
where
\begin{eqnarray}
I_{syn,i} &=& \frac{1}{d_i^{(in)}} \sum_{j=1 (j \ne i)}^N J_{ij} w_{ij} g_j(t) (x_i - X_{syn}), \label{eq:PD4}\\
g_j(t) &=& \sum_{f=1}^{F_j} E(t-t_f^{(j)}-\tau_l); \nonumber \\
E(t) &=& \frac{1}{\tau_d - \tau_r} (e^{-t/\tau_d} - e^{-t/\tau_r}) \Theta(t). \label{eq:PD5}
\end{eqnarray}
Here, the state of the $i$th neuron at a time $t$ (measured in units of milliseconds) is characterized by three state variables: the fast membrane potential $x_i$, the fast recovery current $y_i,$ and the slow adaptation current $z_i$. The parameter values used in our computations are listed in Table \ref{tab:Parm}. More details on external stimulus on the single HR neuron, synaptic currents and synaptic plasticity, and numerical integration of the governing equations are given in the following subsections.

\begin{figure}
\includegraphics[width=0.8\columnwidth]{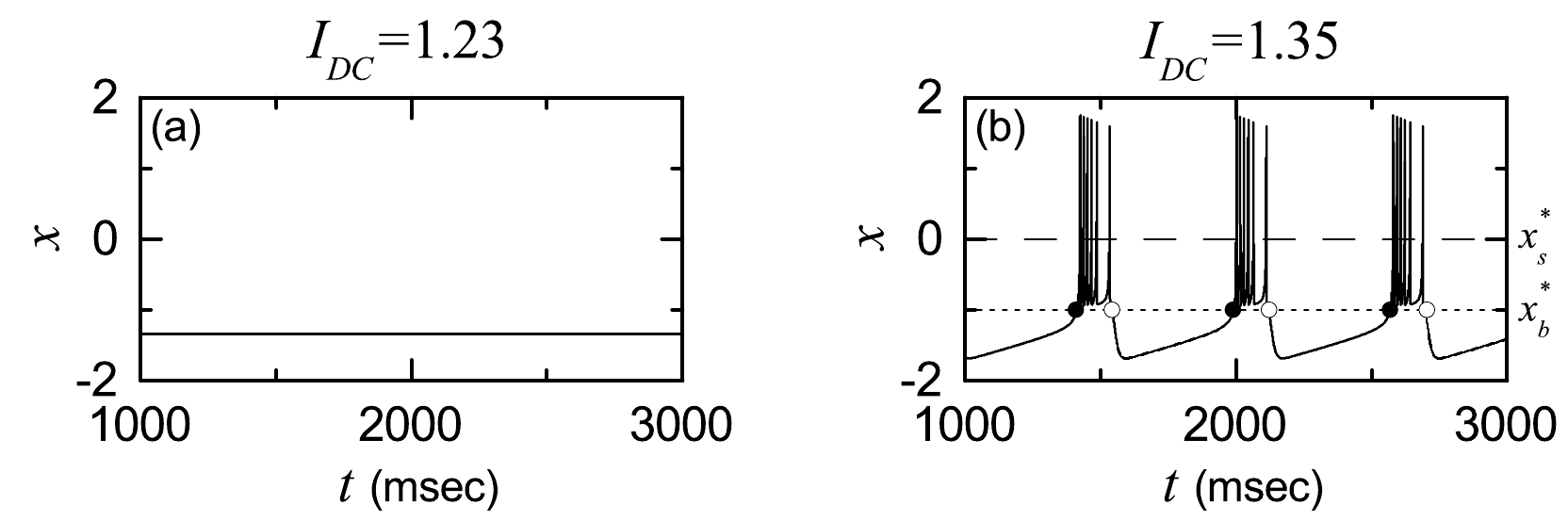}
\caption{Single bursting HR neuron for $D=0$. (a) Time series of  $x(t)$ for the resting state when $I_{DC}=1.23$. (b) Time series of $x(t)$ for the bursting state when $I_{DC}=1.35$.
The dotted horizontal line ($x^*_b=-1$) and the dashed horizontal line ($x^*_s=0$) represent the bursting and spiking thresholds, respectively. The solid and open circles denote the burst onset and offset times, respectively.
}
\label{fig:Single}
\end{figure}

\subsection{External Stimulus to Each HR Neuron}
\label{subsubsec:Sti}
Each bursting HR neuron (whose parameter values are in the 1st item of Table \ref{tab:Parm} \cite{Longtin})  is stimulated by a DC current $I_{DC,i}$ and an independent Gaussian white noise $\xi_i$ [see the 5th and the 6th terms in Eq.~(\ref{eq:PD1})] satisfying $\langle \xi_i(t) \rangle =0$ and $\langle \xi_i(t)~\xi_j(t') \rangle = \delta_{ij}~\delta(t-t')$, where $\langle\cdots\rangle$ denotes the ensemble average. The intensity of noise $\xi_i$ is controlled by the parameter $D$. Figure \ref{fig:Single}(a) shows a resting state of a single HR neuron for $I_{DC}=1.23$ in the absence of noise (i.e., $D=0$).
As $I_{DC}$ passes a threshold $I_{DC}^* (\simeq 1.26)$, each single HR neuron exhibits a transition from a resting state to a bursting state.
For the suprathreshold case of $I_{DC}=1.35$, deterministic bursting occurs when neuronal activity alternates, on a slow time scale $(\simeq 578$ msec), between a silent phase and an active (bursting) phase of fast repetitive spikings, as shown in Fig.~\ref{fig:Single}(b). The dotted horizontal line ($x^*_b=-1$) denotes the bursting threshold (the solid and open circles denote the active phase onset and offset times, respectively), while the dashed horizontal line ($x^*_s=0$) represents the spiking threshold within the active phase. An active phase of the bursting activity begins (ends) at a burst onset (offset) time when the membrane potential $x$ of the bursting HR neuron passes the bursting threshold of $x^*_b=-1$ from below (above). For this case, the HR neuron exhibits bursting activity with the slow bursting frequency $f_b (\simeq 1.7$ Hz) [corresponding to the average inter-burst interval (IBI) ($\simeq 578$ msec)]. Throughout this paper, we consider a suprathreshold case such that the value of $I_{DC,i}$ is chosen via uniform random sampling in the range of [1.3,1.4], as shown in the 2nd item of Table \ref{tab:Parm}.

\subsection{Synaptic Currents and Plasticity}
\label{subsec:Syn}
The last term in Eq.~(\ref{eq:PD1}) represents the synaptic couplings of HR bursting neurons. $I_{syn,i}$ of Eq.~(\ref{eq:PD4}) represents a synaptic current injected into the $i$th neuron, and $X_{syn}$ is the synaptic reversal potential. The synaptic connectivity is given by the connection weight matrix $W$ (=$\{ w_{ij} \}$) where  $w_{ij}=1$ if the bursting neuron $j$ is presynaptic to the bursting neuron $i$; otherwise, $w_{ij}=0$.
Here, the synaptic connection is modeled in terms of the Barab\'{a}si-Albert SFN. Then, the in-degree of the $i$th neuron, $d_i^{(in)}$ (i.e., the number of synaptic inputs to the interneuron $i$) is given by $d_i^{(in)} = \sum_{j=1 (j \ne i)}^N w_{ij}$.
The fraction of open synaptic ion channels at time $t$ is denoted by $g(t)$. The time course of $g_j(t)$ of the $j$th neuron is given by a sum of delayed double-exponential functions $E(t-t_f^{(j)}-\tau_l)$ [see Eq.~(\ref{eq:PD5})], where $\tau_l$ is the synaptic delay, and $t_f^{(j)}$ and $F_j$ are the $f$th spike and the total number of spikes of the $j$th neuron at time $t$, respectively. Here, $E(t)$ [which corresponds to contribution of a presynaptic spike occurring at time $0$ to $g_j(t)$ in the absence of synaptic delay] is controlled by the two synaptic time constants: synaptic rise time $\tau_r$ and decay time $\tau_d$, and $\Theta(t)$ is the Heaviside step function: $\Theta(t)=1$ for $t \geq 0$ and 0 for $t <0$. For the inhibitory GABAergic synapse (involving the $\rm{GABA_A}$ receptors), the values of $\tau_l$, $\tau_r$, $\tau_d$, and $X_{syn}$ are listed in the
3rd item of Table \ref{tab:Parm} \cite{GABA}.

\begin{figure}
\includegraphics[width=0.5\columnwidth]{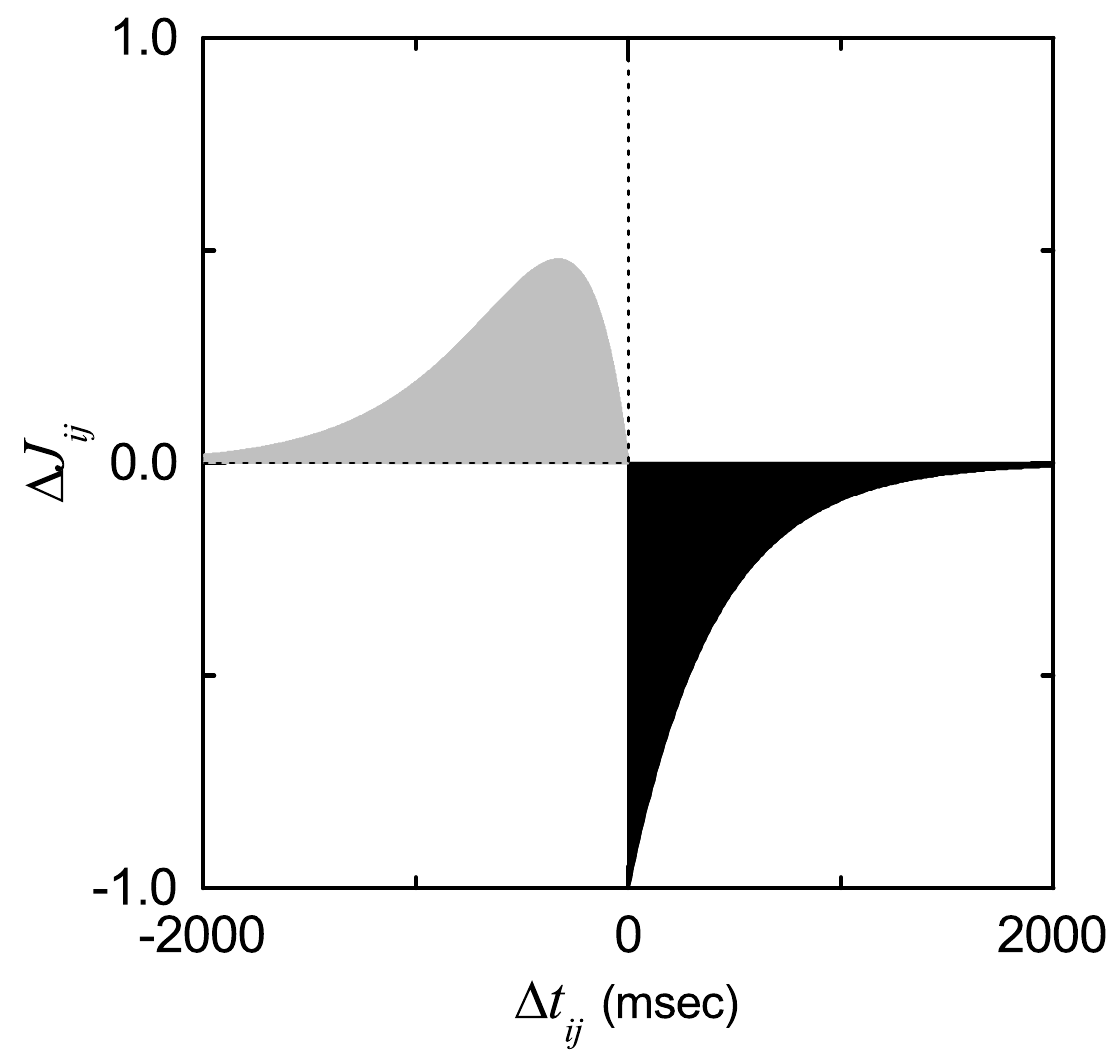}
\caption{Time window for the Anti-Hebbian iSTDP. Plot of synaptic modification $\Delta J_{ij}$ versus $\Delta t_{ij}$ $(=t_i^{(post)} - t_j^{(pre)})$ for $A_+=1$, $A_{-}=1.3$, $\tau_+=410$ msec and $\tau_{-}=330$ msec. $t_i^{(post)}$ and $t_j^{(pre)}$ are burst onset times of the $i$th post-synaptic and the $j$th pre-synaptic neurons, respectively.
}
\label{fig:TW}
\end{figure}

The coupling strength of the synapse from the $j$th pre-synaptic neuron to the $i$th post-synaptic neuron is $J_{ij}$.
Here, we consider a multiplicative iSTDP (dependent on states) for the synaptic strengths $\{ J_{ij} \}$ \cite{Tass2,Multi}.
To avoid unbounded growth and elimination of synaptic connections, we set a range with the upper and the lower bounds: $J_{ij} \in [J_l, J_h]$,
where $J_l=0.0001$ and $J_h=20$. Initial synaptic strengths are normally distributed with the mean $J_0(=12)$ and the standard deviation $\sigma_0(=0.1)$.
With increasing time $t$, the synaptic strength for each synapse is updated with a nearest-burst pair-based STDP rule \cite{SS}:
\begin{equation}
J_{ij} \rightarrow J_{ij} + \delta (J^*-J_{ij})~|\Delta J_{ij}(\Delta t_{ij})|,
\label{eq:MSTDP}
\end{equation}
where $\delta$ $(=0.08)$ is the update rate, $J^*=$ $J_h~(J_l)$ for the LTP (LTD), and $\Delta J_{ij}(\Delta t_{ij})$ is the synaptic modification depending on the relative time difference $\Delta t_{ij}$ $(=t_i^{(post)} - t_j^{(pre)})$ between the nearest burst onset times of the post-synaptic neuron $i$ and the pre-synaptic neuron $j$.
We use an asymmetric anti-Hebbian time window for the synaptic modification $\Delta J_{ij}(\Delta t_{ij})$ \cite{Tass1,Lou,Minesota}:
\begin{equation}
  \Delta J_{ij}(\Delta t_{ij}) = \left\{ \begin{array}{l} -A_{+}~  e^{-\Delta t_{ij} / \tau_{+}} ~{\rm for}~ \Delta t_{ij} > 0\\
  - A_{-}~ \frac{\Delta t_{ij}}{\tau_{-}} ~ e^{\Delta t_{ij} / \tau_{-}} ~{\rm for}~ \Delta t_{ij} \le 0\end{array} \right. ,
\label{eq:TW}
\end{equation}
where $A_+=1.0$, $A_-=1.3$, $\tau_+=410$ msec, and $\tau_-=330$ msec (these values are also given in the 5th item of Table \ref{tab:Parm}).
Figure \ref{fig:TW} shows an asymmetric anti-Hebbian time window for $\Delta J_{ij}(\Delta t_{ij})$ of Eq.~(\ref{eq:TW}) (i.e., plot of $\Delta J_{ij}$ versus $\Delta t_{ij}$).
$\Delta J_{ij}(\Delta t_{ij})$ changes depending on the relative time difference $\Delta t_{ij}$ $(=t_i^{(post)} - t_j^{(pre)})$ between the nearest burst onset times of the post-synaptic neuron $i$ and the pre-synaptic neuron $j$. In contrast to the case of a Hebbian time window for the eSTDP \cite{SSS,SBS,SSSSTDP}, when a post-synaptic burst follows a pre-synaptic burst (i.e., $\Delta t_{ij}$ is positive), LTD of synaptic strength appears
in the black region; otherwise (i.e., $\Delta t_{ij}$ is negative), LTP occurs in the gray region.

\subsection{Numerical Integration}
\label{subsec:NI}
Numerical integration of stochastic differential equations (\ref{eq:PD1})-(\ref{eq:PD3}) is done using the Heun method \cite{SDE} (with the time step $\Delta t=0.01$ msec).
For each realization of the stochastic process, we choose a random initial point $[x_i(0),y_i(0),z_i(0)]$ for the $i$th $(i=1,\dots, N)$ neuron with uniform probability in the range of $x_i(0) \in (-1.5,1.5)$, $y_i(0) \in (-10,0)$, and $z_i(0) \in (1.2,1.5)$.

\section{Effects of Inhibitory STDP on Burst Synchronization}
\label{sec:BS}

\subsection{BS in The Absence of iSTDP}
\label{subsec:NSTDP}
First, we are concerned about the BS (i.e., population synchronization on the slow bursting timescale) in the absence of iSTDP for the case of symmetric attachment with $l_{in}= l_{out}=l^*=15$ in the SFN of $N$ inhibitory Hindmarsh-Rose bursting neurons. The coupling strengths $\{ J_{ij} \}$ are static, and their values are chosen from the Gaussian distribution where the mean $J_0$ is 12 and the standard deviation $\sigma_0$ is 0.1.
We investigate emergence of BS by varying the noise intensity $D$.

\begin{figure}
\includegraphics[width=\columnwidth]{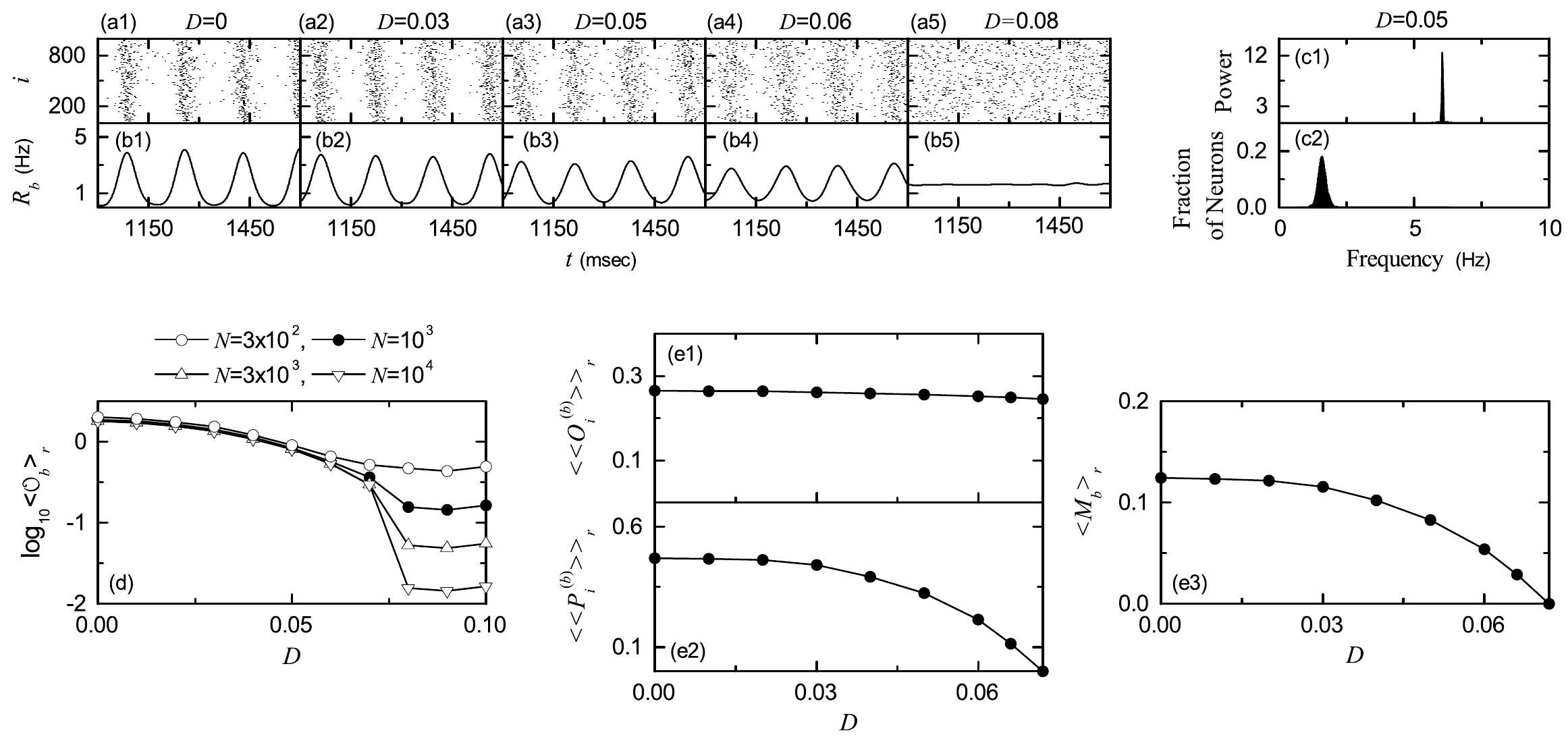}
\caption{Burst synchronization in the absence of iSTDP for the case of symmetric attachment with $l^*=15$; $N=10^3$ except for the case of the bursting order parameter in (d).
Raster plots of burst onset times in (a1)-(a5) and IPBR kernel estimates $R_b(t)$ in (b1)-(b5) for various values of $D=$0, 0.03, 0.05, 0.06, and 0.08. (c1) One-sided power spectrum of $\Delta R_b(t) [=R_b(t) - \overline{R_b(t)}]$ (the overbar represents the time average) with mean-squared amplitude normalization and (c2) distribution of mean bursting rates (MBRs) of individual neurons for $D=0.05$.
(d) Plots of the thermodynamic bursting order parameter $\langle {\cal{O}}_b \rangle_r$ versus $D$. Plots of (e1) the average occupation degree $\langle \langle O_i^{(b)} \rangle \rangle_r$,
(e2) the average pacing degree $\langle \langle P_i^{(b)} \rangle \rangle_r$, and (e3) the statistical-mechanical bursting measure  $\langle M_b \rangle_r$ versus $D$.
}
\label{fig:BS1}
\end{figure}

BS may be well visualized in the raster plot of burst onset times which corresponds to a collection of all trains of burst onset times of individual bursting neurons.
Figures \ref{fig:BS1}(a1)-\ref{fig:BS1}(a5) show such raster plots for various values of $D$.
To see emergence of BS, we employ an (experimentally-obtainable) instantaneous population burst rate (IPBR) which is often used as a collective quantity showing bursting behaviors.
This IPBR may be obtained from the raster plot of burst onset times \cite{Kim1,Kim2,NN-SFN}.
To obtain a smooth IPBR, we employ the kernel density estimation (kernel smoother) \cite{Kernel}. Each burst onset time in the raster plot is convoluted (or blurred) with a kernel function
$K_h(t)$ to obtain a smooth estimate of IPBR $R_b(t)$:
\begin{equation}
R_b(t) = \frac{1}{N} \sum_{i=1}^{N} \sum_{b=1}^{n_i} K_h (t-t_{b}^{(i)}),
\label{eq:IPBR}
\end{equation}
where $t_{b}^{(i)}$ is the $b$th burst onset time of the $i$th neuron, $n_i$ is the total number of burst onset times for the $i$th neuron, and we use a Gaussian
kernel function of band width $h$:
\begin{equation}
K_h (t) = \frac{1}{\sqrt{2\pi}h} e^{-t^2 / 2h^2}, ~~~~ -\infty < t < \infty \label{eq:Gaussian}
\end{equation}
Throughout the paper, the band width $h$ of $K_h(t)$ is 20 msec. Figures \ref{fig:BS1}(b1)-\ref{fig:BS1}(b5) show IPBR kernel estimates $R_b(t)$ for various values of $D$.
For the synchronous case, ``bursting stripes'' (composed of burst onset times and indicating BS) are formed in the raster plot of burst onset times [see Figs.~\ref{fig:BS1}(a1)-\ref{fig:BS1}(a4)], and the corresponding IPBR kernel estimates $R_b(t)$ exhibit oscillations, as shown in Figs.~\ref{fig:BS1}(b1)-\ref{fig:BS1}(b4).
As an example, we consider the case of $D=0.05$. The bursting frequency $f_b$ [i.e., the oscillating frequency of $R_b(t)$] is 6.09 Hz [see Fig.~\ref{fig:BS1}(c1)], while the population-averaged mean bursting rate (MBR)
$\langle f_i^{(b)} \rangle$ of individual bursting neurons is 1.56 Hz [see Fig.~\ref{fig:BS1}(c2)].
For this type of BS, individual bursting neurons fire at lower rates $f_i^{(b)}$ than the bursting frequency $f_b$, and hence only a smaller fraction of bursting neurons fire in each bursting stripe in the raster plot
(i.e., each stripe is sparsely occupied by burst onset times of a smaller fraction of bursting neurons).
In this way, sparse BS occurs, in contrast the full BS where individual neurons show bursting at every global cycle of $R_b(t)$ \cite{SBS}.
On the other hand, in the desynchronized case for $D > D^* (\simeq 0.072)$, burst onset times are completely scattered in the raster plot, and $R_b(t)$ is nearly stationary,
as shown in Figs.~\ref{fig:BS1}(a5) and \ref{fig:BS1}(b5) for $D=0.08$.

Recently, we introduced a realistic bursting order parameter, based on $R_b(t)$, for describing transition from BS to desynchronization \cite{Kim2}.
The mean square deviation of $R_b(t)$,
\begin{equation}
{\cal{O}}_b \equiv \overline{(R_b(t) - \overline{R_b(t)})^2},
 \label{eq:Order}
\end{equation}
plays the role of an order parameter ${\cal{O}}_b$; the overbar represents the time average. This bursting order parameter may be regarded as a thermodynamic measure because it
concerns just the macroscopic IPBR kernel estimate $R_b(t)$ without any consideration between $R_b(t)$ and microscopic individual burst onset times.
In the thermodynamic limit of $N \rightarrow \infty$, the bursting order parameter ${\cal{O}}_b$ approaches a non-zero (zero) limit value for the synchronized (desynchronized) state.
Hence, the bursting order parameter can determine synchronized and desynchronized states.
Figure \ref{fig:BS1}(d) shows plots of $\log_{10} \langle {\cal{O}}_b \rangle_r$ versus $D$.
In each realization, we discard the first time steps of a stochastic trajectory as transients for $10^3$ msec, and then we numerically compute ${\cal{O}}_b$ by following the stochastic trajectory
for $3 \times 10^4$ msec. Hereafter, $\langle \cdots \rangle_r$ denotes an average over 20 realizations.
For $D < D^* (\simeq 0.072)$, the bursting order parameter $\langle {\cal{O}}_b \rangle_r$ approaches a non-zero limit value, and hence BS appears.
On the hand hand, when passing $D^*$ a transition from BS to desynchronization occurs, because $\langle {\cal{O}}_b \rangle_r$ tends to zero with increasing $N$.

In the absence of noise (i.e., $D=0$), sparse bursting stripes (indicating sparse BS) appear successively in the raster plot of burst onset times, and the IPBR kernel estimate $R_b(t)$ exhibits an oscillatory behavior.
However, as $D$ is increased, sparse bursting stripes become smeared gradually, as shown in the cases of $D=0.03,$ 0.05, and 0.06, and hence the amplitudes of $R_b (t)$ decrease in a slow way.
Eventually, as $D$ passes $D^*,$ desynchronization occurs due to overlap of smeared bursting stripes.
Then, burst onset times are completely scattered without forming any bursting stripes, and hence the IPBR kernel estimate $R_b(t)$ becomes nearly stationary, as shown
for the case of $D=0.08$.

We characterize sparse BS in the range of $0 \leq D < D^*$ by employing a statistical-mechanical bursting measure $M_b$ \cite{Kim2}. For the case of BS, bursting stripes appear regularly in the raster plot of burst onset times.
The bursting measure $M^{(b)}_i$ of the $i$th bursting stripe is defined by the product of the occupation degree $O^{(b)}_i$ of burst onset times (denoting the density of the $i$th bursting stripe) and the
pacing degree $P^{(b)}_i$ of burst onset times (representing the smearing of the $i$th bursting stripe):
\begin{equation}
M^{(b)}_i = O^{(b)}_i \cdot P^{(b)}_i.
\label{eq:BMi}
\end{equation}
The occupation degree $O^{(b)}_i$ of burst onset times in the $i$th bursting stripe is given by the fraction of bursting neurons:
\begin{equation}
   O^{(b)}_i = \frac {N_i^{(b)}} {N},
\label{eq:OD}
\end{equation}
where $N_i^{(b)}$ is the number of bursting neurons in the $i$th bursting stripe.
For the case of full BS, all bursting neurons exhibit burstings in each bursting stripe in the raster plot of burst onset times, and hence the occupation degree $O_i^{(b)}$
of Eq.~(\ref{eq:OD}) in each bursting stripe becomes 1. On the other hand, in the case of sparse BS, only some fraction of bursting neurons show burstings in each bursting stripe,
and hence the occupation degree $O_i^{(b)}$ becomes less than 1.
In our case of BS, $O^{(b)}_i <1$ in the range of $ 0 \leq D < D^*$, and hence sparse BS occurs.
The pacing degree $P^{(b)}_i$ of burst onset times in the $i$th bursting stripe can be determined in a statistical-mechanical way by taking into account their contributions to the macroscopic IPBR kernel estimate $R_b(t)$.
Central maxima of $R_b(t)$ between neighboring left and right minima of $R_b(t)$ coincide with centers of bursting stripes in the raster plot. A global cycle starts from a left minimum of
$R_b(t)$, passes a maximum, and ends at a right minimum.
An instantaneous global phase $\Phi^{(b)}(t)$ of $R_b(t)$ was introduced via linear interpolation in the region forming a global cycle
(for details, refer to Eqs.~(14) and (15) in \cite{Kim2}).  Then, the contribution of the $k$th microscopic burst onset time in the $i$th bursting stripe occurring at the time $t_k^{(b)}$ to $R_b(t)$ is
given by $\cos \Phi^{(b)}_k$, where $\Phi^{(b)}_k$ is the global phase at the $k$th burst onset time [i.e., $\Phi^{(b)}_k \equiv \Phi^{(b)}(t_k^{(b)})$]. A microscopic burst onset time makes the most constructive (in-phase)
contribution to $R_b(t)$ when the corresponding global phase $\Phi^{(b)}_k$ is $2 \pi n$ ($n=0,1,2, \dots$), while it makes the most destructive (anti-phase) contribution to $R_b(t)$ when $\Phi^{(b)}_k$
is $2 \pi (n-1/2)$. By averaging the contributions of all microscopic burst onset times in the $i$th bursting stripe to $R_b(t)$, we obtain the pacing degree of burst onset times in the $i$th stripe:
\begin{equation}
 P^{(b)}_i ={ \frac {1} {B_i}} \sum_{k=1}^{B_i} \cos \Phi^{(b)}_k,
\label{eq:PACING}
\end{equation}
where $B_i$ is the total number of microscopic burst onset times in the $i$th stripe.

By averaging $M_i^{(b)}$ of Eq.~(\ref{eq:BMi}) over a sufficiently large number $N_b$ of bursting stripes, we obtain the realistic statistical-mechanical bursting measure $M_b$, based on the IPBR kernel estimate $R_b(t)$:
\begin{equation}
M_b =  {\frac {1} {N_b}} \sum_{i=1}^{N_b} M^{(b)}_i.
\label{eq:BM}
\end{equation}
We follow $3 \times 10^3$ bursting stripes in each realization and get $\langle M_b \rangle_r$ via average over 20 realizations.
Figures \ref{fig:BS1}(e1)-\ref{fig:BS1}(e3) show the average occupation degree $\langle \langle O_i^{(b)} \rangle \rangle_r$, the average pacing degree $\langle \langle P_i^{(b)} \rangle \rangle_r$, and
the statistical-mechanical bursting measure $\langle M_b \rangle_r$, respectively.
With increasing $D$ from 0 to $D^*$, $\langle \langle O_i^{(b)} \rangle \rangle_r$ (denoting the density of bursting stripes in the raster plot) decreases very slowly from 0.27 to 0.25 (i.e.,
$\langle \langle O_i^{(b)} \rangle \rangle_r$ is nearly constant). The average pacing degree $\langle \langle P_i^{(b)} \rangle \rangle_r$ represents well the smearing degree of bursting stripes in the raster plot
 of burst onset times [shown in Figs.~\ref{fig:BS1}(a1)-\ref{fig:BS1}(a4)]. With increasing $D$ from 0 to $D^*$, $\langle \langle P_i^{(b)} \rangle \rangle_r$ decreases to zero smoothly due to complete overlap of sparse bursting stripes. Through product of the average occupation and pacing degrees of burst onset times, the statistical-mechanical bursting measure $\langle M_b \rangle_r$ is obtained.
Since $\langle \langle O_i^{(b)} \rangle \rangle_r$ is nearly constant, $\langle M_b \rangle_r$ behaves like the case of $\langle \langle P_i^{(b)} \rangle \rangle_r$.

\begin{figure}
\includegraphics[width=0.8\columnwidth]{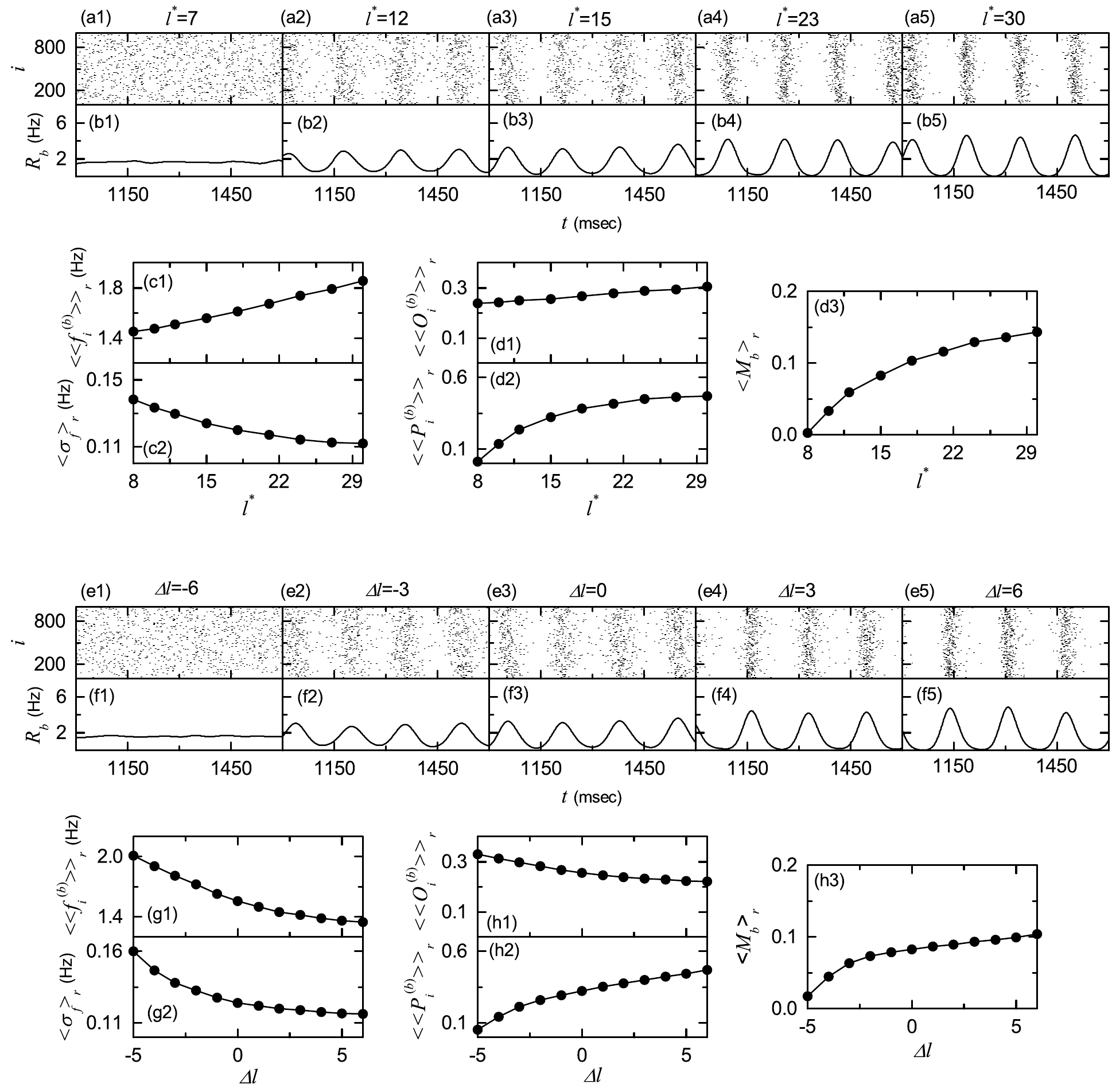}
\caption{Effect of network architecture on BS in the absence of iSTDP for $D=0.05$; $N=10^3$.
Symmetric preferential attachment with $l_{in}=l_{out}=l^*$. Raster plots of burst onset times in (a1)-(a5) and IPBR kernel estimates $R_b(t)$ in (b1)-(b5) for various values of $l^*$.
Plots of (c1) population-averaged MBRs $\langle \langle f_i^{(b)} \rangle \rangle_r$ and (c2) standard deviations $\langle \sigma_f \rangle_r$ from $\langle f_i^{(b)} \rangle$ versus $l^*$.
Plots of (d1) the average occupation degree $\langle \langle O_i^{(b)} \rangle \rangle_r$, (d2) the average pacing degree $\langle \langle P_i^{(b)} \rangle \rangle_r$, and (d3) the statistical-mechanical bursting measure  $\langle M_b \rangle_r$ versus $l^*$.
Asymmetric preferential attachment with $l_{in}= l^* + \Delta l$ and $l_{out} = l^* - \Delta l$ ($l^*=15$).
Raster plots of burst onset times in (e1)-(e5) and IPBR kernel estimates $R_b(t)$ in (f1)-(f5) for various values of $\Delta l$.
Plots of (g1) population-averaged MBRs $\langle \langle f_i^{(b)} \rangle \rangle_r$ and (g2) standard deviations $\langle \sigma_f \rangle_r$ from $\langle f_i^{(b)} \rangle$ versus $\Delta l$.
Plots of (h1) the average occupation degree $\langle \langle O_i^{(b)} \rangle \rangle_r$, (h2) the average pacing degree $\langle \langle P_i^{(b)} \rangle \rangle_r$, and (h3) the statistical-mechanical bursting measure  $\langle M_b \rangle_r$ versus $\Delta l$.
}
\label{fig:BS2}
\end{figure}

We now fix the value of $D$ at $D=0.05$ where BS occurs for the case of symmetric attachment with $l^*=15$ [see Figs.~\ref{fig:BS1}(a3) and \ref{fig:BS1}(b3)], and investigate the effect of scale-free connectivity on BS
by varying the degree of symmetric attachment $l^*$ (i.e., $l_{in}=l_{out}  = l^*$) and the asymmetry parameter $\Delta l$ of asymmetric attachment [i.e., $l_{in}= l^* + \Delta l$ and $l_{out}= l^* - \Delta l$ ($l^*=15$)].

We first consider the case of symmetric attachment, and study  its effect on BS by varying the degree $l^*$. Figures \ref{fig:BS2}(a1)-\ref{fig:BS2}(a5) show the raster plots of burst onset times for various values of $l^*$. Their corresponding IPBR kernel estimates $R_b(t)$ are also shown in Figs.~\ref{fig:BS2}(b1)-\ref{fig:BS2}(b5).
With increasing $l^*$ from $15$ (i.e., the case studied above), bursting stripes in the raster plots of burst onset times become clearer (e.g., see the cases of $l^*=23$ and 30), which also leads to increase in the oscillating amplitudes of $R_b(t)$, in comparison with that for the case of $l^*=15$. In this way, as $l^*$ is increased from 15, the degree of BS becomes better.
On the other hand, with decreasing $l^*$ from 15, bursting stripes become more smeared (e.g., see the case of $l^*=12$), and hence the oscillating amplitude of $R_b(t)$ decreases. Thus, as $l^*$ is decreased from 15, the degree of BS becomes worse. Eventually, the population state becomes desynchronized for $l^* \leq 7$, as shown in Figs.~\ref{fig:BS2}(a1) and \ref{fig:BS2}(b1) where burst onset times are completely scattered and $R_b(t)$ becomes nearly stationary.

Effects of $l^*$ on network topology were characterized in Ref.~\cite{NN-SFN}, where the group properties of the SFN were studied in terms of the average path length $L_p$ and the betweenness centralization $B_c$
by varying $l^*$. The average path length $L_p$ represents typical separation between two nodes in the network, and it characterizes global efficiency of information transfer between distant nodes \cite{BA2,NN-SFN}.
The group betweenness centralization $B_c$ denotes how much the load of communication traffic is concentrated on the head hub (with the highest degree) \cite{BETC1,BETC2,NN-SFN}.
Large $B_c$ implies that load of communication traffic is much concentrated on the head hub, and hence the head hub tends to become overloaded by the communication traffic passing through it
\cite{BC3}. With increasing $l^*$, both the average path length $L_p$ and the betweenness centralization $B_c$ become smaller, due to increase in the total number of connections
(see Figs.~11(c) and 11(e) in \cite{NN-SFN}). Hence, typical separation between neurons becomes shorter, and load of communication traffic concentrated on the head neuron also becomes smaller.
As a result, as $l^*$ is increased, efficiency of global communication between neurons (i.e., global transfer of neural information between neurons via synaptic connections) becomes better,
which may contribute to increase in the degree of BS.

Along with network topology, we also consider individual dynamics which change depending on the synaptic inputs with the in-degree $d^{(in)}$ of Eq.~(\ref{eq:PD5}).
The in-degree distribution affects MBRs of individual bursting neurons (e.g., see Figs.~11(g1)-11(g5) in Ref.~\cite{NN-SFN}).
As $l^*$ is increased, the average in-degree $\langle d^{(in)} \rangle$ (=$ \frac {1} {N} \sum_{i=1}^{N} d^{(in)}_i$) increases, which favors the pacing between bursting neurons.
Consequently, with increasing $l^*$, the population-averaged MBR $\langle \langle f_i^{(b)} \rangle \rangle_r$ increases, and the standard deviation $\langle \sigma_f \rangle_r$ decreases
(i.e., population-averaged individual dynamics become better), as shown in Figs.~\ref{fig:BS2}(c1) and \ref{fig:BS2}(c2), which may also contribute to increase in the degree of BS.
Figures \ref{fig:BS2}(d1)-\ref{fig:BS2}(d3) show the average occupation degree $\langle \langle O_i^{(b)} \rangle \rangle_r$, the average pacing degree $\langle \langle P_i^{(b)} \rangle \rangle_r$, and
the statistical-mechanical bursting measure $\langle M_b \rangle_r$, respectively.
With increasing $l^*$,  $\langle \langle P_i^{(b)} \rangle \rangle_r$ increases markedly due to decrease in the standard deviation $\langle \sigma_f \rangle_r$. On the other hand, $\langle \langle O_i^{(b)} \rangle \rangle_r$ increases a little due to a slight increase in the population-averaged MBR $\langle \langle f_i^{(b)} \rangle \rangle_r$. Then, $\langle M_b \rangle_r$ (given by the product of $\langle \langle O_i^{(b)} \rangle \rangle_r$ and $\langle \langle P_i^{(b)} \rangle \rangle_r$) increases distinctly as in the case of $\langle \langle P_i^{(b)} \rangle \rangle_r$.
Consequently, as $l^*$ is increased from 15, the degree of BS increases due to both better individual dynamics and better efficiency of global communication between nodes (resulting from the increased number of total connections). On the other hand, with decreasing $l^*$ from 15, both individual dynamics and effectiveness of communication between nodes become worse (resulting from the decreased number of total connections),
and hence the degree of BS decreases.

As the second case of network architecture, we consider the case of asymmetric attachment; $l_{in}= l^* + \Delta l$ and $l_{out}= l^* - \Delta l$ ($l^*=15$).
For the case of asymmetric attachment, the total number of inward and outward connections is fixed (i.e., $ l_{in} + l_{out} = 30$ =constant), in contrast to the case of symmetric attachment where with increasing $l^*$ the number of total connections increases. We investigate the effect of asymmetric attachment on BS by varying the asymmetry parameter $\Delta l$.

Figures \ref{fig:BS2}(e1)-\ref{fig:BS2}(e5) show the raster plots of burst onset times for various values of $\Delta l$. Their corresponding IPBR kernel estimates $R_b(t)$ are also shown in Figs.~\ref{fig:BS2}(f1)-\ref{fig:BS2}(f5).
As $\Delta l$ is increased from $0$, bursting stripes in the raster plots of burst onset times become clearer (e.g., see the cases of $\Delta l=3$ and 6), and hence the oscillating amplitudes of $R_b(t)$ become larger than that for the case of $\Delta l=0$. In this way, with increasing $\Delta l$ from 0, the degree of BS becomes better. On the other hand, with decreasing $\Delta l$ from 0, bursting stripes become more smeared (e.g., see the case of $\Delta l = -3$), which results in decrease in the oscillating amplitudes of $R_b(t)$. Hence, as $\Delta l$ is decreased from 0, the degree of BS becomes worse.
Eventually, the population state becomes desynchronized for $\Delta l \leq -6$, as shown in Figs.~\ref{fig:BS2}(e1) and \ref{fig:BS2}(f1) where burst onset times are completely scattered and $R_b(t)$ becomes nearly stationary.

As $| \Delta l |$ (the magnitude of $\Delta l$) is increased, both $L_p$ and $B_c$ increase symmetrically, independently of the sign of $\Delta l$, due to increased mismatching between the in- and the out-degrees
(see Figs.~13(c) and 13(d) in \cite{NN-SFN}). The values of $L_p$ and $B_c$ for both cases of different signs but the same magnitude (i.e., $\Delta l$ and $- \Delta l$) become the same because both inward and
outward connections are involved equally in computations of $L_p$ and $B_c$. Due to the effects of $\Delta l$ on $L_p$ and $B_c$, with increasing $| \Delta l |$, efficiency of global communication between nodes becomes
worse, independently of the sign of $\Delta l$, which may contribute to decrease in the degree of BS.

However, individual dynamics change depending on the sign of $\Delta l$ because of different average in-degrees $\langle d^{(in)} \rangle$.
Particularly, the distribution of MBRs of individual bursting neurons vary depending on $\Delta l$ (e.g., see Figs.~13(g1)-13(g5) in Ref.~\cite{NN-SFN}).
As $\Delta l$ is increased from 0, $\langle d^{(in)} \rangle$ increases, which tends to favor the pacing between bursting neurons.
Hence, the standard deviation $\langle \sigma_f \rangle_r$ (for the distribution of MBRs $\{ f_i^{(b)} \}$) decreases, as shown in Fig.~\ref{fig:BS2}(g2).
In addition, with increasing $\Delta l$ from 0, the population-averaged MBR $\langle \langle f_i^{(b)} \rangle \rangle_r$ also decreases, because of
increase in average inhibition given to individual neurons (resulting from increased population-averaged in-degrees) [see Fig.~\ref{fig:BS2}(g1)].
In contrast, with decreasing $\Delta l$ from 0, the average in-degree $\langle d^{(in)} \rangle$ decreases, which tends to disfavor the pacing between bursting neurons.
Therefore, the standard deviation $\langle \sigma_f \rangle_r$ increases [see Fig.~\ref{fig:BS2}(g2)]. Moreover, as $\Delta l$ is decreased from 0, the
population-averaged MBR $\langle \langle f_i^{(b)} \rangle \rangle_r$ also increases, due to decrease in average inhibition given to individual neurons (resulting from
decreased population-averaged in-degrees) [see Fig.~\ref{fig:BS2}(g1)].
In this way, as $\Delta l$ is increased (decreased) from 0, individual dynamics become better (worse), which may contribute to increase (decrease) in the degree of BS.
Figures \ref{fig:BS2}(h1)-\ref{fig:BS2}(h3) show the average occupation degree $\langle \langle O_i^{(b)} \rangle \rangle_r$, the average pacing degree $\langle \langle P_i^{(b)} \rangle \rangle_r$, and
the statistical-mechanical bursting measure $\langle M_b \rangle_r$, respectively.
With increasing $\Delta l$,  $\langle \langle P_i^{(b)} \rangle \rangle_r$ increases distinctly, mainly due to decrease in the standard deviation $\langle \sigma_f \rangle_r$ (which overcomes worse efficiency of communication).
In contrast, as $\Delta l$ is increased, $\langle \langle O_i^{(b)} \rangle \rangle_r$ decreases a little due to a slight decrease in the population-averaged MBR $\langle \langle f_i^{(b)} \rangle \rangle_r$.
However,  $\langle \langle P_i^{(b)} \rangle \rangle_r$ increases more rapidly than decrease in $\langle \langle O_i^{(b)} \rangle \rangle_r$.
Hence, with increasing $\Delta l$ $\langle M_b \rangle_r$ (given by the product of $\langle \langle O_i^{(b)} \rangle \rangle_r$ and $\langle \langle P_i^{(b)} \rangle \rangle_r$) also increases.
Consequently, as $\Delta l$ is decreased from 0, the degree of BS decreases because both individual dynamics and efficiency of communication between nodes are worse. On the other hand, with increasing
$\Delta l$ from 0, the degree of BS increases mainly because of better individual dynamics overcoming worse efficiency of communication.

\begin{figure}
\includegraphics[width=\columnwidth]{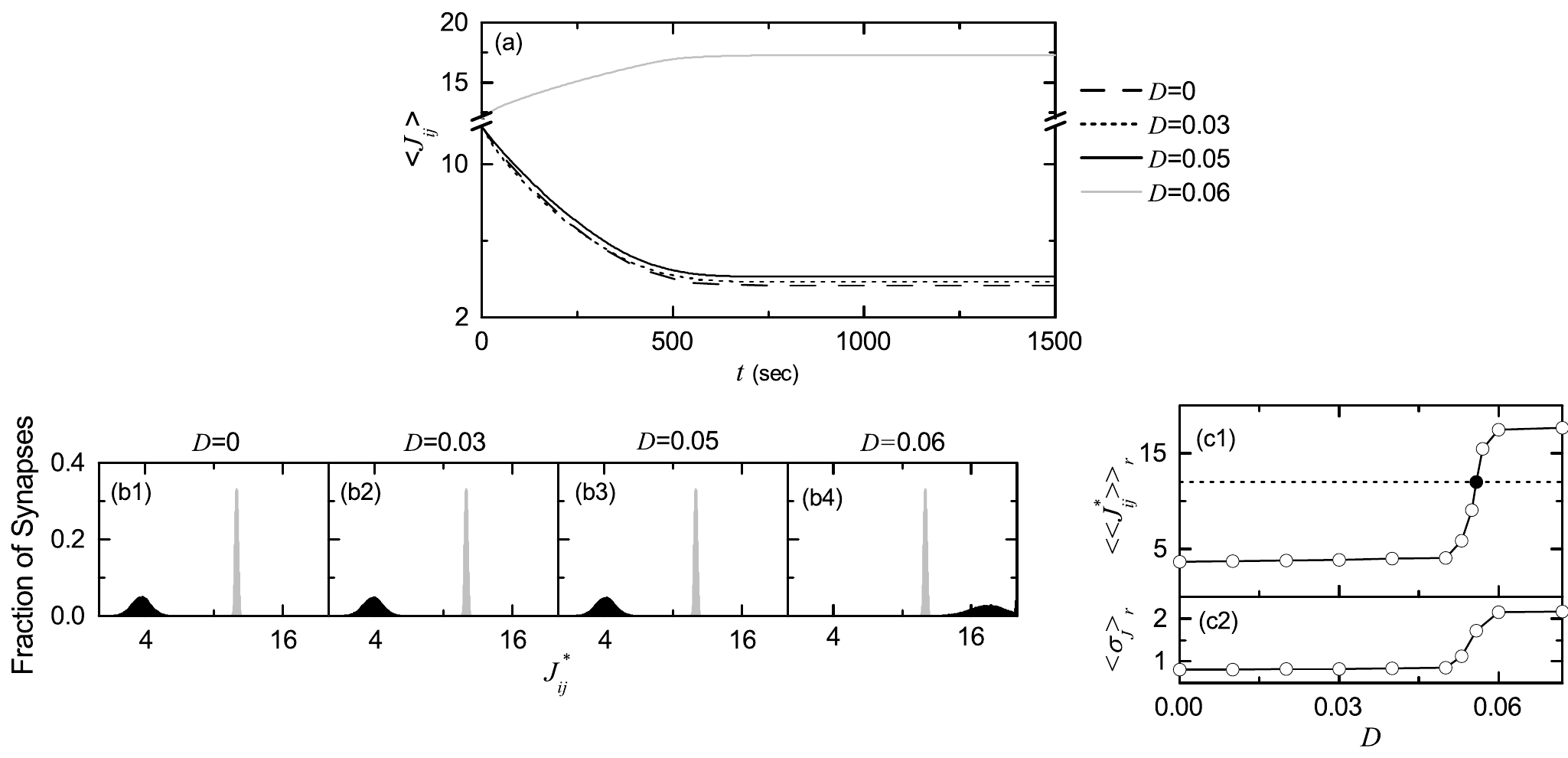}
\caption{Effect of iSTDP on BS for the case of symmetric attachment with $l^*=15$; $N=10^3$.
(a) Time-evolutions of population-averaged synaptic strengths $\langle J_{ij} \rangle$ for various values of $D$.
(b1)-(b4) Histograms for the fraction of synapses versus $J^*_{ij}$ (saturated limit values of $J_{ij}$  at 1000 sec) are shown in black color for various values of $D$; for comparison,
initial distributions of synaptic strengths $\{ J_{ij} \}$  are also shown in gray color. Plots of (c1) population-averaged limit values of synaptic strengths  $\langle \langle J^*_{ij} \rangle \rangle_r$
and (c2) standard deviations $\langle \sigma_J \rangle_r$ versus $D$.
}
\label{fig:STDP1}
\end{figure}

\subsection{Effects of iSTDP on BS}
\label{sec:iSTDP}
In this subsection, we study the effect of iSTDP on BS [occurring for $0 \leq D < D^* (\simeq 0.072)$ in the absence of iSTDP].
The initial values of synaptic strengths $\{ J_{ij} \}$ are chosen from the Gaussian distribution with the mean $J_0$ (=12) and the standard
deviation $\sigma_0$ (=0.1). Here, we employ an anti-Hebbian time window for the synaptic modification $\Delta J_{ij}(\Delta t_{ij})$ of Eq.~(\ref{eq:TW}).
Then, $J_{ij}$ for each synapse is updated according to a nearest-burst pair-based STDP rule of Eq.~(\ref{eq:MSTDP}).

We first consider the case of symmetric attachment with $l^*=15$, and investigate the effect of iSTDP on BS by varying $D$.
Figure \ref{fig:STDP1}(a) shows time-evolutions of population-averaged synaptic strengths $\langle J_{ij} \rangle$ for various values of $D$; $\langle \cdots \rangle$ represents an
average over all synapses. For each case of $D=0,$ 0.03, and 0.05, $\langle J_{ij} \rangle$ decreases monotonically below its initial value $J_0$ (=12), and it approaches a saturated limit value $\langle
J_{ij}^* \rangle$ nearly at $t=1000$ sec. Consequently, LTD occurs for these values of $D$.
On the other hand, for $D=0.06$ $\langle J_{ij} \rangle$ increases monotonically above $J_0$, and approaches a saturated limit value $\langle J_{ij}^* \rangle$. As a result, LTP occurs for the case of $D=0.06$.
Histograms for fraction of synapses versus $J_{ij}^*$ (saturated limit values of $J_{ij}$ at $t=1000$ sec) are shown in black color for various values of $D$ in Figs.~\ref{fig:STDP1}(b1)-\ref{fig:STDP1}(b4); the bin size for each histogram is 0.1. For comparison, initial distributions of synaptic strengths $\{ J_{ij} \}$ (i.e., Gaussian distributions whose mean $J_0$ and standard deviation $\sigma_0$ are 12 and 0.1, respectively) are also shown in gray color. For the cases of LTD ($D=0,$ 0.03, and 0.05), their black histograms lie on the left side of the initial gray histograms, and hence their population-averaged values $\langle J_{ij}^*
\rangle$ become smaller than the initial value $J_0$. On the other hand, the black histogram for the case of LTP ($D=0.06$) is shifted to the right side of the initial gray
histogram, and hence its population-averaged value $\langle J_{ij}^* \rangle$ becomes larger than $J_0$.
For both cases of LTD and LTP, their black histograms are wider than the initial gray histograms [i.e., the standard deviations $\sigma_J$ are larger than the initial one $\sigma_0$].
Figure \ref{fig:STDP1}(c1) shows a plot of population-averaged limit values of synaptic strengths $\langle \langle J_{ij}^* \rangle \rangle_r$ versus $D$. Here, the horizontal dotted line represents the initial average value of coupling strengths $J_0$, and the threshold value $D_{th}$ $(\simeq 0.0558)$ for LTD/LTP (where $\langle \langle J_{ij}^* \rangle \rangle_r = J_0$) is represented by a solid circle. Hence, LTD occurs in a range of BS
($0 \leq D < D_{th}$); BS in the absence of iSTDP appears in the range of $ 0 \leq D < D^*$.
As $D$ is decreased from $D_{th}$, $\langle \langle J_{ij}^* \rangle \rangle_r$ decreases monotonically.
In contrast, LTP takes place in a smaller range of BS (i.e., $D_{th} < D < D^*$), and
with increasing $D$ from $D_{th}$ $\langle \langle J_{ij}^* \rangle \rangle_r$ increases monotonically.
Figure \ref{fig:STDP1}(c2) shows plots of standard deviations $\langle \sigma_J \rangle_r$ versus $D$.
With increasing $D$ from 0 to $D^*$, $\langle \sigma_J \rangle_r$ increases, and all the values of $\langle \sigma_J \rangle_r$ are larger than the initial value $\sigma_0$ (=0.1).

\begin{figure}
\includegraphics[width=0.75\columnwidth]{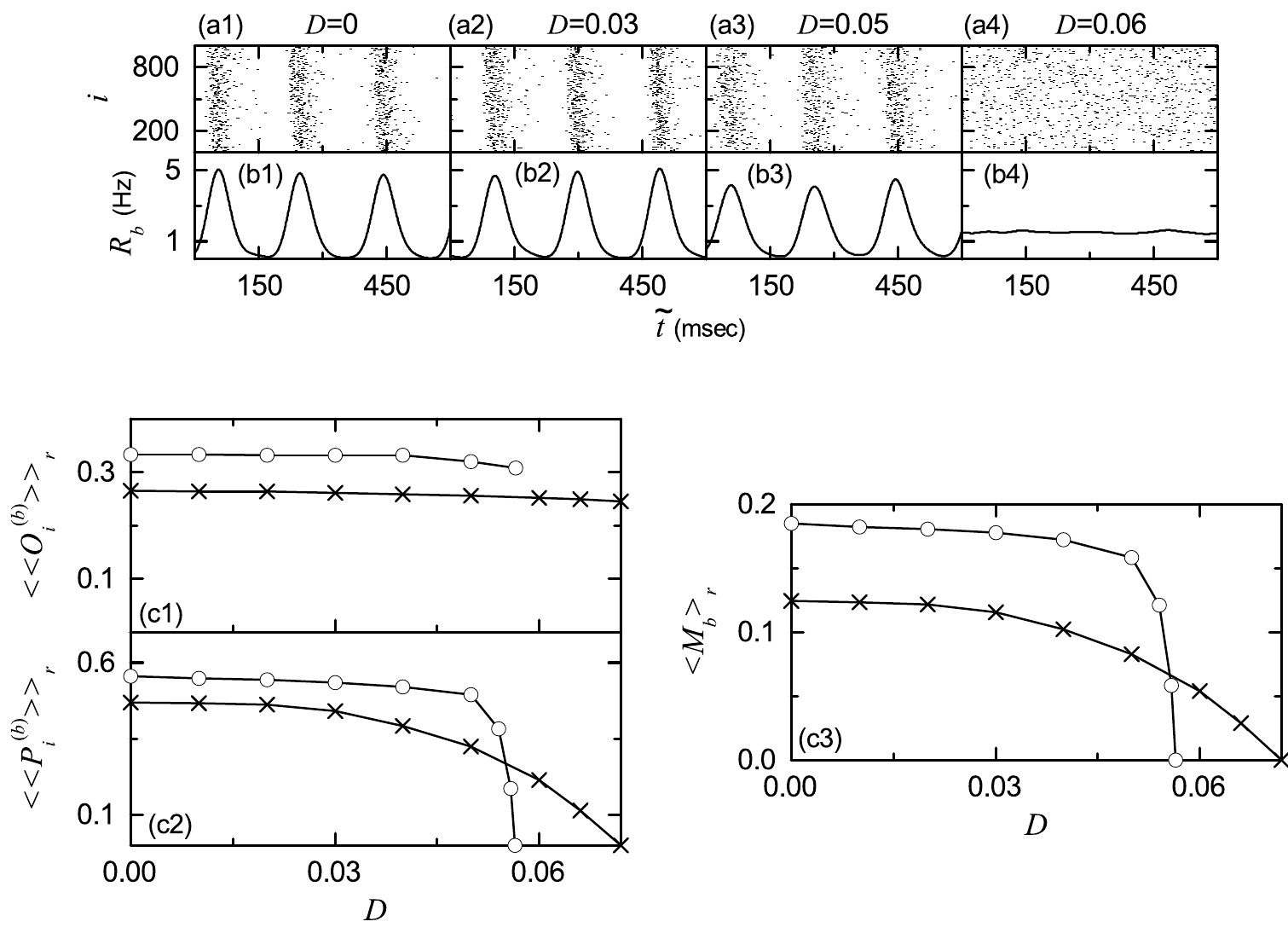}
\caption{Characterization of BS after the saturation time ($t=1000$ sec) for the case of symmetric attachment with $l^*=15$; $N=10^3$.
Raster plots of burst onset times in (a1)-(a4) and IPBR kernel estimates $R_b(t)$ in (b1)-(b4) for various values of $D$ after the saturation time, where
$t=t^*$ (saturation time = 1000 sec) + $\widetilde{t}$.
Plots of (c1) the average occupation degree $\langle \langle O_i^{(b)} \rangle \rangle_r$ (open circles), (c2) the average pacing degree $\langle \langle P_i^{(b)} \rangle \rangle_r$ (open circles), and (c3) the statistical-mechanical bursting measure $\langle M_b \rangle_r$ (open circles) versus $D$. For comparison, $\langle \langle O_i^{(b)} \rangle \rangle_r$, $\langle \langle P_i^{(b)} \rangle \rangle_r$,
and $\langle M_b \rangle_r$ versus $D$ in the absence of iSTDP are also denoted by crosses.
}
\label{fig:STDP2}
\end{figure}

The LTD (LTP) has a tendency to increase (decrease) the degree of BS because of decrease (increase) in the mean value of synaptic inhibition strengths, and
the increased standard deviations tend to decrease the degree of BS. The effects of LTD and LTP on BS after the saturation time ($t=1000$ sec) may be well shown in the raster plot of burst onset times and the corresponding IPBR kernel estimate $R_b(t)$, which are given in Figs.~\ref{fig:STDP2}(a1)-\ref{fig:STDP2}(a4) and Figs.~\ref{fig:STDP2}(b1)-\ref{fig:STDP2}(b4) for various values of $D$, respectively.
When compared with Figs.~\ref{fig:BS1}(a1)-\ref{fig:BS1}(a3) and Figs.~\ref{fig:BS1}(b1)-\ref{fig:BS1}(b3) in the absence of iSTDP,
the degrees of BS for the case of $D=0,$ 0.03, and 0.05 are increased [i.e., the amplitudes of $R_b(t)$ are increased] due to dominant effect of LTD (overcoming the effect of increased standard deviation).
On the other hand, in the case of $D=0.06$ the population state becomes desynchronized [i.e., $R_b(t)$ becomes nearly stationary] due to the effects of both LTP and increased standard deviation
[compare Figs.~\ref{fig:STDP2}(a4) and \ref{fig:STDP2}(b4) with Figs.~\ref{fig:BS1}(a4) and \ref{fig:BS1}(b4)].
Due to inhibition, the roles of LTD and LTP in inhibitory synaptic plasticity are reversed in comparison with those in excitatory synaptic plasticity where the degree of population synchronization is increased (decreased) via LTP (LTD) \cite{SSS,SBS,SSSSTDP}.

In the presence of iSTDP, we also characterize population behaviors for BS after the saturation time ($t=1000$ sec) in the range of $0 \leq D < D^{**} (\simeq 0.0565)$ (where BS persists in the presence of iSTDP).
For comparison, corresponding quantities for BS in the absence of iSTDP are also shown in the range of  $0 \leq D < D^* (\simeq 0.072)$ (where BS appears in the absence of iSTDP).
Figures \ref{fig:STDP2}(c1) and \ref{fig:STDP2}(c2) show the average occupation degree $\langle \langle O_i^{(b)} \rangle \rangle_r$ and the average pacing degree $\langle \langle P_i^{(b)} \rangle \rangle_r$ (represented by open circles), respectively; for comparison, $\langle \langle O_i^{(b)} \rangle \rangle_r$  and $\langle \langle P_i^{(b)} \rangle \rangle_r$ (denoted by crosses) are also shown in the case without iSTDP.
In the region of $0 \leq D < D_{th} (\simeq 0.0558)$, the values of $\langle \langle O_i^{(b)} \rangle \rangle_r$ (open circles) are larger than those (crosses) in the absence of iSTDP, due to LTD (decreased mean synaptic inhibition). In most region of LTD, the values of $\langle \langle P_i^{(b)} \rangle \rangle_r$ (open circles) are also larger than those (crosses) in the absence of iSTDP, because of dominant effect of LTD (overcoming the
effect of increased standard deviation). In the region of $D_{th} < D < D^{**}$, $\langle \langle O_i^{(b)} \rangle \rangle_r$ (open circles) decreases just a little.
However, for the case of $\langle \langle P_i^{(b)} \rangle \rangle_r$ (open circles), a rapid transition to the case of $\langle \langle P_i^{(b)} \rangle \rangle_r =0$ occurs due to the effects of both LTP and increased standard deviation, in contrast to the smooth decrease in $\langle \langle P_i^{(b)} \rangle \rangle_r$ (crosses) in the absence of iSTDP.
The statistical-mechanical bursting measure $\langle M_b \rangle_r$ (combining the effect of both the average occupation and pacing degrees) is shown in open circles in Fig.~\ref{fig:STDP2}(c3).
Behaviors of $\langle M_b \rangle_r$  are similar to those of $\langle \langle P_i^{(b)} \rangle \rangle_r$, because the values of $\langle \langle O_i^{(b)} \rangle \rangle_r$ are nearly constant.
A Matthew effect is found to occur in the presence of iSTDP. In most range of $D$ with LTD, good BS (with higher bursting measure) gets better since the effect of mean LTD is dominant in comparison to the effect of increased standard deviation. On the other hand, in the range of $D$ with LTP, bad BS (with lower bursting measure) gets worse because of the effects of both LTP and increased standard deviation.
Thus, near the threshold $D_{th}$ a rapid transition to desynchronization (i.e. the case of $\langle M_b \rangle_r=0$) occurs via LTP, in contrast to a smooth transition in the absence of iSTDP.

\begin{figure}
\includegraphics[width=\columnwidth]{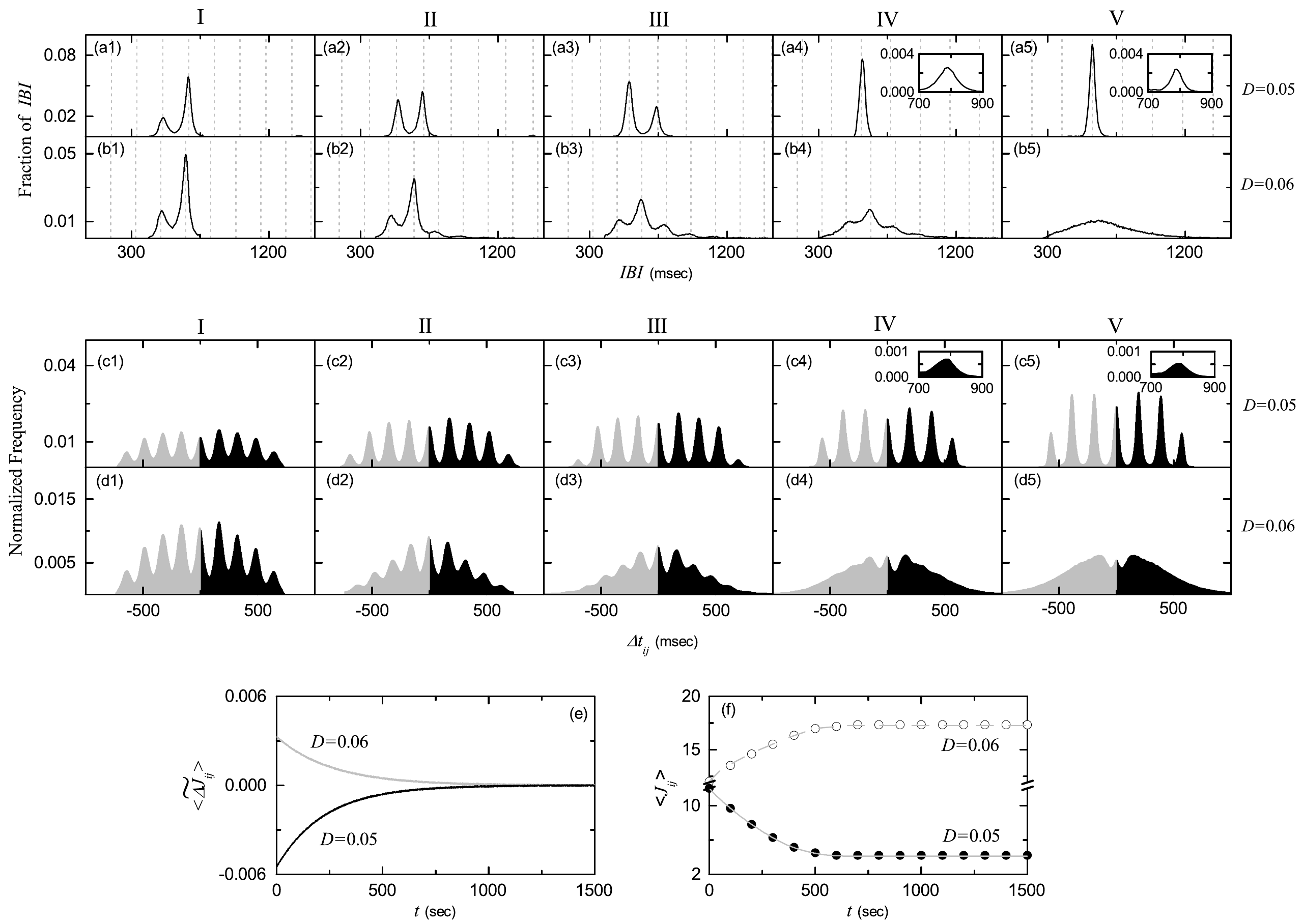}
\caption{Microscopic investigations on emergences of LTD and LTP for the case of symmetric attachment with $l^*=15$; $N=10^3$.
Time-evolutions of the IBI histograms for $D = 0.05$ in (a1)-(a5) and $D = 0.06$ in (b1)-(b5); 5 stages are shown in I (starting from 0 sec), II (starting from 100 sec), III (starting from 250 sec), IV (starting from 500 sec), and V (starting from 800 sec). Vertical dotted lines represent multiples of the global period $T_G$ of the IPBR $R_b(t)$.
Time-evolutions of the normalized histogram $H(\Delta t_{ij})$ for the distributions of time delays $\{ \Delta t_{ij} \}$ between the pre- and the post-synaptic burst onset times for $D = 0.05$ in (c1)-(c5) and $D = 0.06$ in (d1)-(d5); 5 stages are shown in I (starting from 0 sec), II (starting from 100 sec), III (starting from 250 sec), IV (starting from 500 sec), and V (starting from 800 sec).
Time-evolutions of (e) multiplicative synaptic modification $\langle {\widetilde{\Delta J_{ij}}} \rangle$ and (f) population-averaged synaptic strength $\langle J_{ij} \rangle$ (obtained by an approximate method);
gray solid and dashed lines represent ones (obtained by direct calculations) for $D=$ 0.05 and 0.06 in Fig.~\ref{fig:STDP1}(a), respectively.
}
\label{fig:STDP3}
\end{figure}

From now on, we make an intensive investigation on emergences of LTD and LTP of synaptic strengths via a microscopic method based on the distributions of time delays $\{ \Delta t_{ij} \}$ between the pre- and
the post-synaptic burst onset times. For understanding time-evolution of the distribution of $\{ \Delta t_{ij} \}$, we first study the time-evolutions of the IBI histograms for $D=0.05$ (LTD) and 0.06 (LTP), which are shown
in Figs.~\ref{fig:STDP3}(a1)-\ref{fig:STDP3}(a5) and \ref{fig:STDP3}(b1)-\ref{fig:STDP3}(b5), respectively. Here, we consider 5 stages, represented by I (starting from 0 sec), II (starting from 100 sec), III (starting from 250 sec), IV (starting from 500 sec), and  V (starting from 800 sec). For each stage, we get the IBI histogram from IBIs obtained from all bursting neurons during 3 sec, and the bin size is 2.5 msec.
For the case of $D=0.05$ (LTD), the IBI histogram at the stage I consists of two peaks; the 1st lower peak and the 2nd higher peak, located at $3~T_G$ and $4~T_G$ [$T_G$ (denoted by the vertical dotted lines): global period of the IPBR $R_b(t)$], respectively. Hence, individual neurons exhibit burstings intermittently at every 3rd or 4th global cycle of $R_b(t)$. Since the amplitude of the 2nd peak is larger, occurrence of burstings at every 4th global cycle is more probable. However, as the time $t$ is increased (i.e., with increase in the level of stage) the amplitude of the 1st peak increases, while that of the 2nd peak is reduced.
In the stages IV and V, extremely small 2nd peaks are shown in the insets. Hence, eventually individual neurons show burstings nearly at every 3rd global cycle of $R_b(t)$.
In the case of $D=0.06$ (LTP), the IBI histogram at the stage I is also composed of the 1st lower peak and the 2nd higher peak, located at $3~T_G$ and $4~T_G$, respectively. However, with increasing the stage, peaks become
wider, and merging between them occurs. Eventually, at the stage V a single broad peak appears. These distributions of IBIs for $D=0.05$ and 0.06 affect the distribution of $\{ \Delta t_{ij} \}$, as shown below.

Figures \ref{fig:STDP3}(c1)-\ref{fig:STDP3}(c5) and \ref{fig:STDP3}(d1)-\ref{fig:STDP3}(d5) show time-evolutions of normalized histograms $H(\Delta t_{ij})$ for the distributions of time delays $\{ \Delta t_{ij} \}$ for $D=0.05$ and 0.06, respectively; the bin size in each histogram is 2.5 msec. Like the above case, we also consider 5 stages, represented by I (starting from 0 sec), II (starting from 100 sec), III (starting from 250 sec), IV (starting from 500 sec), and  V (starting from 800 sec). At each stage, we get the distribution of $\{ \Delta t_{ij} \}$ for all synaptic pairs during 0.5 sec and obtain the normalized histogram by dividing the distribution with the total number of synapses. Here, LTD and LTP occur in the black ($\Delta t>0$) and the gray ($\Delta t <0$) parts, respectively.
For $D=0.05$ (LTD), at the stage I, 9 peaks [the central 1st-order peak and each pair of left and right higher $i$th-order peaks ($i$=2, 3, 4, and 5)] appear in each histogram of $\{ \Delta t_{ij} \}$, in contrast to the case
of full BS where just 3 peaks (the main central peak and a pair of minor left and right peaks)
appear (see Figs. 8(c1)-8(c6) in \cite{SBS}). Due to sparse burstings at every 3rd or 4th global cycle of $R_b(t)$, nearest-neighboring pre- and post-synaptic burst onset times may appear in the following separate stripes in the raster plot of burst onset times, such as the nearest-neighboring, the next-nearest-neighboring, the next-next-nearest-neighboring, and the next-next-next-nearest-neighboring stripes, as well as in the same stripe. As a result, 9 peaks appear in the distribution of $H(\Delta t_{ij})$. For the case of stage I, the right black part (LTD) is dominant, in comparison with the left gray part (LTP), and hence the overall net LTD begins to emerge. As the stage is increased, the peaks from the 1st to the 4th order become intensified (i.e., peaks become narrowed, and then they become sharper), because the amplitude of the 1st peak at $3~T_G$ in the IBI histogram is getting dominantly increased. On the other hand, the last 5th-order peaks become very small (see the insets in the stages IV and V where the right 5th-order peak is shown), since the amplitude of the 2nd peak at $4~T_G$ in the IBI histogram  is getting very small. At the stage V, the effect of LTD in the black part tends to nearly cancel out the effect of LTP in the gray part.
For the case of $D=0.06$ (LTP), at the 1st stage, 9 peaks also appear in the histogram of $\{ \Delta t_{ij} \}$, as in the case of $D=0.05$. In this case of stage I, the left gray part (LTP) is dominant, in comparison with the right black part (LTD), and hence the overall net LTP begins to occur. However, as the level of stage is increased, peaks become wider, and merging tendency between the peaks is intensified. At the stage V, only one broad central peak seems to appear.
At the stage V, the effect of LTP in the gray part tends to nearly cancel out the effect of LTD in the black part.

We now consider successive time intervals $I_k \equiv (t_{k},t_{k+1})$, where $t_k=0.5 \cdot (k-1)$ sec ($k=1,2,3,\dots$).
As the time $t$ is increased, we get the $k$th normalized histogram $H_k(\Delta t_{ij})$ ($k=1,2,3,\dots$)
in each $k$th time interval $I_k$, through the distribution of $\{ \Delta t_{ij} \}$ for all synaptic pairs during 0.5 sec.
Then, from Eq.~(\ref{eq:MSTDP}), we get the population-averaged synaptic strength $\langle J_{ij} \rangle_k$ recursively:
\begin{equation}
\langle J_{ij} \rangle_{k} = \langle J_{ij} \rangle_{k-1} + \delta \cdot \langle \widetilde{\Delta J_{ij}}(\Delta t_{ij})  \rangle_{k},
\label{eq:ASS1}
\end{equation}
where $\langle J_{ij} \rangle_0=J_0$ (=12) and $\langle \cdots \rangle_k$ means the average over the distribution of time delays $\{ \Delta t_{ij} \}$ for all synaptic pairs in the $k$th
time interval. Here, the multiplicative synaptic modification $\widetilde{\Delta J_{ij}}(\Delta t_{ij})$ is given
by the product of the multiplicative factor ($J^*-J_{ij}$) [$J_{ij}:$ synaptic coupling strength at the $(k-1)$th stage] and the absolute value of synaptic modification
$| \Delta J_{ij}(\Delta t_{ij}) |$:
\begin{equation}
 \widetilde{\Delta J_{ij}}(\Delta t_{ij})  =  (J^* - J_{ij})~ |\Delta J_{ij}(\Delta t_{ij})|.
\label{eq:ASS2}
\end{equation}
Then, we get the population-averaged multiplicative synaptic modification $\langle \widetilde{\Delta J_{ij}}(\Delta t_{ij}) \rangle_{k}$ at the $k$th stage
through a population-average approximation where $J_{ij}$ is replaced by its population average $\langle J_{ij} \rangle_{k-1}$ at the $(k-1)$th stage:
\begin{equation}
 \langle \widetilde{\Delta J_{ij}}(\Delta t_{ij}) \rangle_k  \simeq (J^*- \langle J_{ij} \rangle_{k-1})~ \langle |\Delta J_{ij}(\Delta t_{ij})| \rangle_k.
\label{eq:ASS3}
\end{equation}
Here, $\langle |\Delta J_{ij}(\Delta t_{ij})| \rangle_k$ can be easily obtained from the $k$th normalized histogram $H_k(\Delta t_{ij})$:
\begin{equation}
  \langle |\Delta J_{ij}(\Delta t_{ij})| \rangle_{k}  \simeq  \sum_{\rm bins} H_{k} (\Delta t_{ij}) \cdot | \Delta J_{ij} (\Delta t_{ij}) |.
\label{eq:ASS4}
\end{equation}
By using Eqs.~(\ref{eq:ASS1}), (\ref{eq:ASS3}), and (\ref{eq:ASS4}), we get approximate values of $\langle \widetilde{\Delta J_{ij}} \rangle_k$ and $\langle J_{ij} \rangle_{k}$ in a recursive way.
Figure \ref{fig:STDP3}(e) shows time-evolutions of $\langle \widetilde{\Delta J_{ij}} \rangle$ for $D=0.05$ (black curve) and $D=0.06$ (gray curve).
$\langle \widetilde{\Delta J_{ij}} \rangle$ for $D=0.05$ is negative. On the other hand, $\langle \widetilde{\Delta J_{ij}} \rangle$ for $D=0.06$ is positive.
For both cases they converge toward nearly zero near the saturation time ($t=1000$ sec) since the effects of LTD and LTP in the normalized histograms are nearly cancelled out.
The time-evolutions of $\langle J_{ij} \rangle$ for $D=0.05$ (solid circles) and $D=0.06$ (open circles) are also given in Fig.~\ref{fig:STDP3}(f).
The approximately-obtained values for $\langle J_{ij} \rangle$ are found to agree well with directly-obtained ones [denoted by the gray solid (dashed) line for $D=0.05$ (0.06)] in Fig.~\ref{fig:STDP1}(a).
In this way, LTD (LTP) emerges for $D=0.05$ (0.06).

\begin{figure}
\includegraphics[width=0.75\columnwidth]{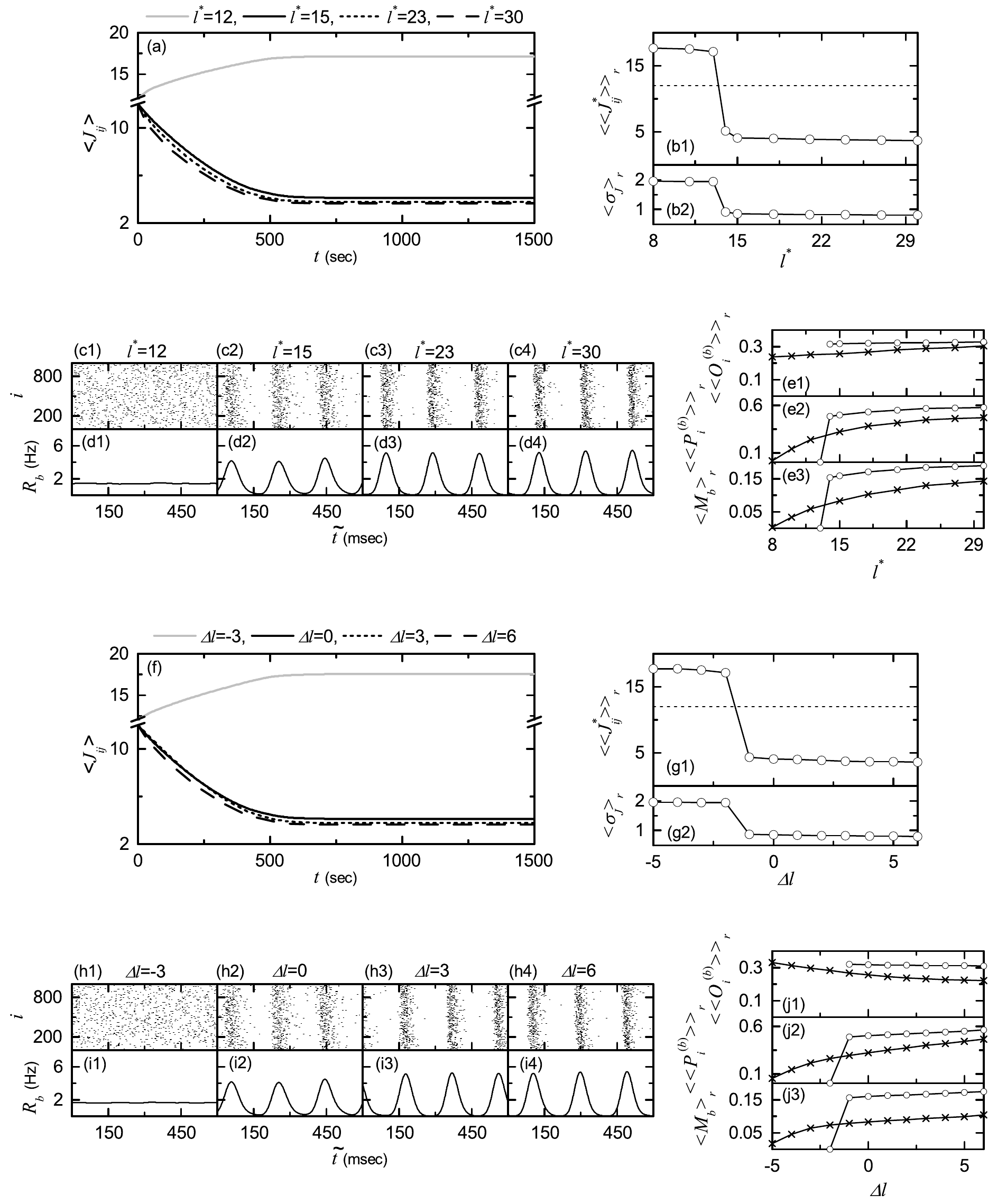}
\caption{Effect of network architecture on BS in the presence of iSTDP for $D=0.05$; $N=10^3$.
Symmetric preferential attachment with $l_{in} = l_{out} = l^*$. (a) Time-evolutions of population-averaged synaptic strengths $\langle J_{ij} \rangle$ for various values of $l^*$.
Plots of (b1) population-averaged limit values of synaptic strengths $\langle \langle J^*_{ij} \rangle \rangle_r$ ($J^*_{ij}:$ saturated limit values of $J_{ij}$ at $t$= 1000 msec)
and (b2) standard deviations $\langle \sigma_J \rangle_r$ versus $l^*$.
Raster plots of burst onset times in (c1)-(c4) and IPBR kernel estimates $R_b(t)$ in (d1)-(d4) for various values of $l^*$ after the saturation time, where $t=t^*$ (saturation time: 1000 sec) + $\widetilde{t}$.
Plots of (e1) the average occupation degree $\langle \langle O_i^{(b)} \rangle \rangle_r$ (open circles), (e2) the average pacing degree $\langle \langle P_i^{(b)} \rangle \rangle_r$ (open circles), and (e3) the statistical-mechanical bursting measure $\langle M_b \rangle_r$ (open circles) versus $l^*$. For comparison, $\langle \langle O_i^{(b)} \rangle \rangle_r$, $\langle \langle P_i^{(b)} \rangle \rangle_r$,
and $\langle M_b \rangle_r$ versus $l^*$ in the absence of iSTDP are also denoted by crosses.
Asymmetric preferential attachment with $l_{in}= l^* + \Delta l$ and $l_{out} = l^* - \Delta l$ ($l^*=15$)
(f) Time-evolutions of population-averaged synaptic strengths $\langle J_{ij} \rangle$ for various values of $\Delta l$.
Plots of (g1) population-averaged limit values of synaptic strengths $\langle \langle J^*_{ij} \rangle \rangle_r$ ($J^*_{ij}:$ saturated limit values of $J_{ij}$ at $t$= 1000 msec)
and (g2) standard deviations $\langle \sigma_J \rangle_r$ versus $\Delta l$.
Raster plots of burst onset times in (h1)-(h4) and IPBR kernel estimates $R_b(t)$ in (i1)-(i4) for various values of $\Delta l$ after the saturation time, where $t=t^*$ (saturation time: 1000 sec) + $\widetilde{t}$.
Plots of (j1) the average occupation degree $\langle \langle O_i^{(b)} \rangle \rangle_r$ (open circles), (j2) the average pacing degree $\langle \langle P_i^{(b)} \rangle \rangle_r$ (open circles), and (j3) the statistical-mechanical bursting measure $\langle M_b \rangle_r$ (open circles) versus $\Delta l$. For comparison, $\langle \langle O_i^{(b)} \rangle \rangle_r$, $\langle \langle P_i^{(b)} \rangle \rangle_r$,
and $\langle M_b \rangle_r$ versus $\Delta l$ in the absence of iSTDP are also denoted by crosses.
}
\label{fig:STDP4}
\end{figure}

Finally, in the presence of iSTDP, we investigate the effect of network architecture on BS for $D=0.05$ by varying the symmetric attachment degree $l^*$ and the asymmetry parameter $\Delta l$.
We first consider the case of symmetric attachment (i.e., $l_{in}=l_{out}  = l^*$).
Figure \ref{fig:STDP4}(a) shows time-evolutions of population-averaged synaptic strengths $\langle J_{ij} \rangle$ for various values of $l^*$.
For each case of $l^*=15,$ 23, and 30, $\langle J_{ij} \rangle$ decreases monotonically below its initial value $J_0$ (=12), and it approaches  a saturated limit value $\langle
J_{ij}^* \rangle$ nearly at $t=1000$ sec. As a result, LTD occurs for these values of $l^*$.
On the other hand, for $l^*=12$ $\langle J_{ij} \rangle$ increases monotonically above $J_0$, and  converges toward a saturated limit value $\langle J_{ij}^* \rangle$.
As a result, for this case LTP takes place.
Figure \ref{fig:STDP4}(b1) shows a plot of population-averaged limit values of synaptic strengths $\langle \langle J_{ij}^* \rangle \rangle_r$ versus $l^*$; the horizontal dotted line represents the initial average value of coupling strengths $J_0$ (= 12). For $l^* \geq 14$ LTD occurs, while for $l^* \leq 13$ LTP takes place.
Figure \ref{fig:STDP4}(b2) also shows plots of standard deviations $\langle \sigma_J \rangle_r$ versus $l^*$.
With increasing $l^*$, $\langle \sigma_J \rangle_r$ decreases, but all the values of $\langle \sigma_J \rangle_r$ are larger than the initial value $\sigma_0$ (=0.1)
The LTD (LTP) tends to increase (decrease) the degree of BS due to decrease (increase) in the mean value of synaptic inhibition strengths, and
increased standard deviations have a tendency to decrease the degree of BS.
We consider the effects of LTD/LTP on BS after the saturation time $t^*$ (= 1000 sec). Figures \ref{fig:STDP4}(c1)-\ref{fig:STDP4}(c4) and Figures \ref{fig:STDP4}(d1)-\ref{fig:STDP4}(d4) show raster plots of burst onset times and the corresponding IPBR kernel estimates $R_b(t)$ for various values of $l^*$, respectively.
Due to the dominant effect of LTD (overcoming the effect of increased standard deviation), the degrees of BS for the case of $l^* =15,$ 23, and 30 are increased so much when compared with Figs.~\ref{fig:BS2}(a3)-\ref{fig:BS2}(a5) and Figs.~\ref{fig:BS2}(b3)-\ref{fig:BS2}(b5) in the absence of iSTDP. In contrast, for the case of $l^*=12$ the population states become desynchronized because of the effects of LTP and increased standard deviation.

We also characterize population behaviors for the BS in terms of the average occupation degree $\langle \langle O_i^{(b)} \rangle \rangle_r$, the average pacing degree $\langle \langle P_i^{(b)} \rangle \rangle_r$, and the statistical-mechanical bursting measure $\langle M_b \rangle_r$. Figure \ref{fig:STDP4}(e1) and \ref{fig:STDP4}(e2) show plots of $\langle \langle O_i^{(b)} \rangle \rangle_r$ and $\langle \langle P_i^{(b)} \rangle \rangle_r$ (denoted by open circles) versus $l^*$, respectively; for  comparison, $\langle \langle O_i^{(b)} \rangle \rangle_r$ and $\langle \langle P_i^{(b)} \rangle \rangle_r$ in the absence of iSTDP are also shown in crosses. For $l^* \geq 14$, the values of $\langle \langle O_i^{(b)} \rangle \rangle_r$ and $\langle \langle P_i^{(b)} \rangle \rangle_r$ (open circles) are larger than those (crosses) in the absence of iSTDP, because of dominant effect of LTD (overcoming the effect of increased standard deviation). However, in the region of $l^* \leq 13$, a rapid transition to desynchronization (i.e. the case of $\langle \langle P_i^{(b)} \rangle \rangle_r =0$) occurs due to the effects of LTP and increased standard deviation, in contrast to the smooth decrease in $\langle \langle P_i^{(b)} \rangle \rangle_r$ (crosses) in the absence of iSTDP.
The statistical-mechanical bursting measure $\langle M_b \rangle_r$ (combining the effect of both the average occupation and pacing degrees) is shown in open circles in Fig.~\ref{fig:STDP4}(e3);
for comparison, $\langle M_b \rangle_r$ in the absence of iSTDP is also shown in crosses. As in the case in Fig.~\ref{fig:STDP2}(c3), a Matthew effect in inhibitory synaptic plasticity occurs.
For $l^* \geq 14$, good BS with higher $M_b$ gets better because the effect of LTD is dominant in comparison with the effect of increased standard deviation).
In contrast, for $l^* \leq 13$, bad BS with lower $M_b$ gets worse via the effects of both LTP and increased standard deviation. Accordingly, a rapid step-like transition to desynchronization occurs, in contrast
to the relatively smooth transition in the absence of iSTDP.

Next, we also consider the case of asymmetric attachment [i.e., $l_{in}= l^* + \Delta l$ and $l_{out}= l^* - \Delta l$ ($l^*=15$)].
Time-evolutions of population-averaged synaptic strengths $\langle J_{ij} \rangle$ for various values of $\Delta l$ are shown
in Fig.~\ref{fig:STDP4}(f). In each case of $\Delta l=0,$ 3, and 6, $\langle J_{ij} \rangle$ decreases monotonically below its initial value $J_0$ (=12), and it converges toward  a saturated limit value $\langle
J_{ij}^* \rangle$ nearly at $t=1000$ sec. Consequently, LTD occurs for these values of $\Delta l$.
In contrast, for $\Delta l=-3,$ $\langle J_{ij} \rangle$ increases monotonically above $J_0$, and
approaches a saturated limit value $\langle J_{ij}^* \rangle$. As a result, for this case LTP takes place.
A plot of population-averaged limit values of synaptic strengths $\langle \langle J_{ij}^* \rangle \rangle_r$ versus $\Delta l$ is shown in Figure \ref{fig:STDP4}(g1);
the horizontal dotted line represents the initial average value of coupling strengths $J_0$ (=12).
For $\Delta l \geq -1$ LTD occurs, while for $\Delta l \leq -2$ LTP takes place.
Plots of standard deviations $\langle \sigma_J \rangle_r$ versus $\Delta l$ are also shown in Fig.~\ref{fig:STDP4}(g2).
As $\Delta l$ is increased, $\langle \sigma_J \rangle_r$ decreases, but all these values of $\langle \sigma_J \rangle_r$ are larger than the initial value $\sigma_0$ (=0.1).
We also consider the effects of LTD (increasing the degree of BS), LTP (decreasing the degree of BS), and increased standard deviation (decreasing the degree of BS) on BS after the saturation time $t^*$ (= 1000 sec).
Figures \ref{fig:STDP4}(h1)-\ref{fig:STDP4}(h4) and Figures \ref{fig:STDP4}(i1)-\ref{fig:STDP4}(i4) show raster plots of burst onset times and the corresponding IPBR kernel estimates $R_b(t)$ for various values of $\Delta l$, respectively. Due to the dominant effect of LTD (overcoming the effect of increased standard deviation) the degrees of BS for the case of $\Delta l =0,$ 3, and 6 are increased much when compared with Figs.~\ref{fig:BS2}(e3)-\ref{fig:BS2}(e5) and Figs.~\ref{fig:BS2}(f3)-\ref{fig:BS2}(f5) in the absence of iSTDP.
On the other hand, in the case of $\Delta l=-3$ the population state becomes desynchronized because of the effects of both LTP and increased standard deviation.

We characterize population behaviors for the BS in terms of the average occupation degree $\langle \langle O_i^{(b)} \rangle \rangle_r$, the average pacing degree $\langle \langle P_i^{(b)} \rangle \rangle_r$, and the statistical-mechanical bursting measure $\langle M_b \rangle_r$. Plots of $\langle \langle O_i^{(b)} \rangle \rangle_r$ and $\langle \langle P_i^{(b)} \rangle \rangle_r$ (denoted by open circles) versus $\Delta l$ are shown in Figs.~\ref{fig:STDP4}(j1) and \ref{fig:STDP4}(j2), respectively; for  comparison, $\langle \langle O_i^{(b)} \rangle \rangle_r$ and $\langle \langle P_i^{(b)} \rangle \rangle_r$ in the absence of iSTDP are also shown in crosses.
For $\Delta l \geq -1$, the values of $\langle \langle O_i^{(b)} \rangle \rangle_r$ and $\langle \langle P_i^{(b)} \rangle \rangle_r$ (open circles) are larger than those (crosses) in the absence of iSTDP, due to the dominant
effect of LTD (overcoming the effect of increased standard deviation). However, in the region of $\Delta l \leq -2$, a rapid step-like transition to the case of $\langle \langle P_i^{(b)} \rangle \rangle_r =0$ takes place because of the effects of both LTP and increased standard deviation, in contrast to the smooth decrease in $\langle \langle P_i^{(b)} \rangle \rangle_r$ (crosses) in the absence of iSTDP.
Figure \ref{fig:STDP4}(j3) shows the statistical-mechanical bursting measure $\langle M_b \rangle_r$ (combining the effect of both the average occupation and pacing degrees) in open circles;
for comparison, $\langle M_b \rangle_r$ in the absence of iSTDP is shown in crosses. Like the above case in Fig.~\ref{fig:STDP4}(e3), a Matthew effect in inhibitory synaptic plasticity occurs.
For $\Delta l \geq -1$, good BS with higher $M_b$ gets better via the dominant effect of LTD (overcoming the effect of increased standard deviation),
while  for $\Delta l \leq -2$ bad BS with lower $M_b$ gets worse via the effects of LTP and increased standard deviation. Consequently, a rapid transition to desynchronization occurs, in contrast to the relatively smooth transition in the absence of iSTDP.

\section{Summary}
\label{sec:SUM}
We are concerned about BS, associated with neural information processes in health and disease, in the Barab\'{a}si-Albert SFN of inhibitory bursting Hindmarsh-Rose neurons. In previous works on BS, inhibitory synaptic plasticity was not considered (i.e., synaptic inhibition strengths were static). On the other hand, in the present work, adaptive dynamics of synaptic inhibition strengths are governed by the iSTDP. An anti-Hebbian time window has been used for the iSTDP update rule, in contrast to the Hebbian time window for the case of eSTDP. The effects of iSTDP on BS have been investigated by varying the noise intensity $D$ for the case of symmetric preferential attachment with $l^*=15$.

Due to inhibition, the roles of LTD (increasing the degree of BS) and LTP (decreasing the degree of BS) for the case of iSTDP are reversed in comparison with those in excitatory synaptic plasticity where the degree of population synchronization is increased (decreased) via LTP (LTD). Increased standard deviations for both cases of LTD and LTP tend to decrease the degree of BS. A Matthew effect has been found in inhibitory synaptic plasticity. In most region of LTD, good BS (with higher bursting measure $M_b$) gets better due to the dominant effect of LTD (overcoming the effect of increased standard deviation).
In contrast, in the region of LTP bad BS (with lower $M_b$) gets worse because of the effects of both LTP and increased standard deviation.
Consequently, near the threshold $D_{th}$ a rapid transition from BS to desynchronization occurs via LTP, in contrast to the relatively smooth transition in the absence of iSTDP.

Emergences of LTD and LTP of synaptic inhibition strengths were investigated via a microscopic method based on the distributions of time delays $\{ \Delta t_{ij} \}$ between the nearest burst onset times of the pre- and the post-synaptic neurons. Time evolutions of normalized histograms $H(\Delta t_{ij})$ were followed for both cases of LTD and LTP. For the case of LTD with $D=0.05$, 9 peaks appear in $H(\Delta t_{ij})$ due to sparse BS, in contrast to the case of full BS where only 3 peaks appear. On the other hand, in the case of LTP with $D=0.06$ merging of such multiple peaks occurs. Based on the normalized histogram at each stage, we recursively obtained population-averaged synaptic inhibition strength $\langle J_{ij} \rangle$ at successive stages by using an approximate recurrence relation. These approximate values of $\langle J_{ij} \rangle$ were found to agree well with directly-calculated ones. In this way, one can understand clearly how microscopic distributions of $\{ \Delta t_{ij} \}$ contribute to $\langle J_{ij} \rangle$.

Futhermore, in the presence of iSTDP, we have also studied the effect of network architecture on BS for a fixed value of $D=0.05$ by varying the symmetric attachment degree $l^*$ and the asymmetry parameter $\Delta l$.
As in the above case of variation in $D$ for $l^*=15$, Matthew effects have also been found to occur for both cases of variations in $l^*$ and $\Delta l$.
For $l^* \geq 14$ and $\Delta l \geq -1$, good BS with higher bursting measure $M_b$ gets better because the effect of LTD is dominant in comparison with the effect of increased standard deviation.
On the other hand, for $l^* \leq 13$ and $\Delta l \leq -2$, bad BS with lower bursting measure $M_b$ gets worse via the effects of both LTP and increased standard deviation.

Finally, we discuss limitations of our present work and future works.
In our present work, we consider only a scale-free complex network. As a future work, it would be interesting to study the effect of iSTDP on BS in other networks with different topology (e.g., small-world or all-to-all networks). In our previous work \cite{FSSiSTDP}, we studied the effect of iSTDP on fast sparse synchronization (FSS) in the small-world neuronal network of inhibitory fast spiking interneurons, and found the same kind of Matthew effect in inhibitory synaptic plasticity; good FSS gets better via LTD, while bad FSS gets worse via LTP. Hence, for our present case of BS, the same kind of Matthew effect in iSTDP is also expected to occur in neuronal networks with different topology.
In the real brain, structural synaptic plasticity (i.e. disappearance, appearance, or rewiring of synapses) also occurs \cite{StPlasticity1,StPlasticity2,StPlasticity3,StPlasticity4,StPlasticity5,StPlasticity6,StPlasticity7,StPlasticity8}, in addition to the case of functional synaptic plasticity where only synaptic strengths change without any structural changes. In our present work, we do not consider this kind of structural synaptic plasticity. Hence, the study on effect of structural plasticity on BS would be interesting as a future work.
In our work, we consider inhibitory synaptic plasticity in the network consisting of just inhibitory bursting neurons. We note that the iSTDP rule may be applicable to spiking neurons as well as bursting neurons, because a spike may be regarded as a burst with a single spike. Hence, as a future work, it would be interesting to study the effect of iSTDP on population synchronization in the network composed of both spiking and bursting neurons.

\section*{Acknowledgments}
This research was supported by the Basic Science Research Program through the National Research Foundation of Korea (NRF) funded by the Ministry of Education
(Grant No. 20162007688).

\end{document}